\documentclass[10pt]{article}
\usepackage{latexsym}  % for Symbol fonts
\usepackage{amssymb}
\usepackage{graphicx}
\usepackage{amsmath}

\begin{document}
%\hfill  2019.07.16
\vspace{5mm}

%%%%%%%%%%%%%%%%%%%%%%%%%%%%%%%

\begin{center}
%{\Large\bf  Detection of extensive cosmic ray air showers by measuring radio emission }
{\large\bf  Detection of extensive cosmic ray air showers by measuring radio emission } 

\vspace{10mm}
{\bf Y. Kawashima}

 {\it Research Center for Nuclear Physics (RCNP)\\
 10-1 Mihogaoka, Ibaraki, Osaka, 567-0047, Japan \\
\it E-mail address: kawa@rncp.osaka-u.ac.jp  \\
         or    bnlkawa@gmail.com}
\end{center}

\begin{quotation}
So far, cosmic ray air showers have been detected using scintillation counter arrays on the ground widely. And also air Cherenkov detection method, which is limited its observation period in moonless nights, has been adopted. The detection method of radio emission from cosmic ray air showers is not new, but rather old method. Radio emission from cosmic ray air showers has not been detected with the method of self-trigger system. If the detection method of radio emission were available, there is no limit of the observation like Cherenkov counter. The developments of high radio frequency (RF) technology might make the detection of radio emission from extensive air showers possible. Antennas calibrated in laboratory are available. With those antennas, we can directly obtain the absolute intensity of radio frequency from cosmic ray air showers. Main issue to detect radio emission is that signal level is quite low. The thermal noise particularly causes background noise source. Those issues and detection method are discussed. We also describe an antenna detection method to generate self-trigger signal and a method to identify air shower events from natural and artificial noises.
\end{quotation}
%%%%%%%%%%%%%%%%%%%%%%%%%%%%%%%%%%%%%%%%%%%%%%%%%%%%
%%%%%%
\vspace{2mm}
%%   \eqno(1.1)
%%{\large\bf 1. \ Introduction }\\ 
\section{Introduction}
%\vspace{2mm}
The first attempt regarding air show detection using antennas for radio emission dates back to 1960s. As long as we know, Linsley \cite{Linsley} began the detection of radio emission from air shower. Next person is Jelley \cite{Jelley}. In 1970s, several persons attempted, one can read papers that were submitted by Allan  \cite{Allan1, Allan2, Allan3}, Atrashkevich  \cite{Atrashkevich} and Mandolesi  \cite{Mandolesi}. However, no one as mentioned above has not directly detected radio emission signal from air shower by the trigger signal generated from radio emission. When referring a trigger signal obtained from scintillation counter arrays, they have observed radio emission signals. A balloon experiment recently reported to have detected air shower events using self-triggered method  \cite{Hoover}. 
 
 In the 21st century, Falcke and Gorham discussed the detection of radio emission from air showers \cite{Falcke}. Their discussion recalled the possibility of detection for radio emission again. In actual, air shower particle detector array began the installation of antenna array to detect radio emission. The CODALEMA \cite{Ardouin} and LOPES \cite{Petrovic} reported their experiments. In particular, CODALEMA shows both spectra detected by the ground particle detector array and antenna array. Their results seem to be very promise for the detection of air shower events by using the method of radio emission.
 
 On the other hand, simulations relating to radio emission from air shower also have been done and continued by Huege et al.  \cite{Huege}. Their simulation results contain so much information to development for detection. In particular, they showed absolute values for electromagnetic field strength from air shower for the first time. They also show the method to identify air shower events by measuring the polarization of radio emission.   Basing on their simulation, we evaluate that it could be possible to detect radio emission from air shower events and also self-triggering method might be realized. Under those considerations, we found that the technology, which handles radio frequency (RF) to accelerate charged particles in accelerator science, was useful to detect radio emission from air showers. Emitted frequencies from air showers might be detected directly using antennas. We describe the method for detection and identification of air shower events from natural or artificial noises. If we could detect air shower events using antennas, the merit of antennas is expected to save the number of detectors and installation cost. If small number of antenna could cover wide area, we could save the cost for detector. Thus we need to estimate the threshold level for the detection of radio emission and show the concrete value of natural background RF power level, so we call thermal noise. We discuss the possibility of detection for radio emission with quantitative analysis in this paper.
 %%%%%%%%
  
%% Section 2
%%{\large\bf 2. \ Radio emission from charged particles under homogeneous magnetic field \\ }
%%\vspace{2mm}
\section{Radio emission from charged particles under homogeneous magnetic field}
%% subsection 2.1
\subsection{\bf{Synchrotron radiation theory}}
When a relativistic charged particle passes through magnetic field, the particle emits energy, which is called synchrotron radiation  \cite {Schwinger, Sokolov}. Even in case of cosmic ray air showers, a high-energy proton or nucleus enters into atmosphere and collides with nitrogen or oxygen atoms. Consequently many charged and neutral particles are yielded through mainly strong and electromagnetic interactions. The earth's magnetic field bends charged particles and synchrotron radiation is produced. Energy emitted from charged particles with relativistic high energy is expressed by a formula   \cite {Wiedermann},

% equation (1)
\begin{equation}
\frac{dP}{dt}=\frac{1}{6\pi \varepsilon _{0}}\frac{e^{2}a^{2}}{c^{3}}\gamma ^{^{4}},
\end{equation}
\[a=\frac{v^{^{2}}}{\rho } ,\]
\[ \gamma =\frac{E}{mc^{^{2}}}=\frac{1}{\sqrt{1-(\frac{v}{c})^{^{2}}}}, \]
\hspace{3cm} $\varepsilon_{0}$ :  dielectric permittivity,\\
\hspace*{3cm} e : charge,\\
\hspace*{3cm} c : speed of light (=2.9979E8 [m/s]),\\
\hspace*{3cm} v : speed of a charged particle [m/s],\\
\hspace*{3cm} $\rho$ : orbital radius [m].\\ 
\hspace*{3cm} m: mass of a charged particle,\\
\hspace*{3cm}E: energy of a charged particle.\\

The equation (1) is rewritten with another parameters,\\
% equation (2)
\begin{equation}
\frac{dP}{dt}=(\frac{2}{3})\gamma_{0}(\frac{{m_{e}}c^{2}}{c^{^{3}}})(\frac{v^{2}}{\rho })^{2}(\frac{E}{mc^{2}})^{4},
\end{equation}
\hspace{3cm} \( r_{0}\) :  classical electron radius ( \(=2.8179  \times 10^{-15}  \) [m] ) \\
\hspace*{3cm} \(m_{e}c^{2} \) : electron mass (  \( =0.511  \times 10^{-3}  \) [GeV]). \\

We consider two cases: one is for {\em electron } and the other for {\em heavier particle } than electron. We need to introduce a famous formula that gives a relation between orbital radius of a charged particle and magnetic field.

% equation (3)
\begin{equation}
\rho [m]=3.336\frac{E[GeV]}{B_{T}[Tesla]}
\end{equation}
\hspace{3cm} \( B_{T}\) :  magnetic field in unit of Tesla. \\

First of all, we deduce synchrotron radiation loss formula for an electron. We input \( r_{0} \) , v=c,  \( mc^{2}=m_{e}c^{2} \) and (3) into (2), and obtain synchrotron radiation loss for an electron,
% equation (4)
\begin{equation}
\frac{dP}{dt}=(\frac{2}{3})r_{0}( \frac{m_{e}c^{2}} {c^{3}})(\frac{B_{T}}{3.336\times E})^{2}c^{4}(\frac{E}{mc^{2}})^{4}
=3.793 \times 10^{11}\cdot B_{T}^{2} \cdot E^{2} [GeV/s] 
=4.22 \times 10^{12}\cdot \frac{E^{4}}{\rho ^{2}} [eV/s].
\end{equation}

To obtain a numerical value for radiation loss, one should use the unit of GeV with respect to E, and meter for $\rho$. Thus we obtained synchrotron radiation loss formula for an electron. 

On the other hand, we can obtain a formula like (4) for heavier charged particles than electron mass, 
% equation (5)
\begin{equation}
\frac{dP}{dt}=(2.8779 \times 10^{-1})(\frac{1}{mc^{2}})^{4}(\frac{E^{4}}{\rho ^{2}}) [eV/s].
\end{equation}
In order to calculate synchrotron radiation loss, we obtained two formulae, which are applicable for electron and heavier particles. 

%% subsection 2.2
\subsection{\bf Energy estimation from radio emission }
A proton with high energy of more than \(10^{16}\) eV enters into atmosphere and interacts with nitrogen or oxygen nucleus. Many pions and kaons are mainly produced. Muons and electrons finally reach the ground. Muons that are produced from decay chain from charged pions and kaons have a long lifetime of 2.2 $\mu$s, so that muons are alive even on the ground. On the other hand, electrons and positrons are yielded through cascade shower processes. Let us consider that a muon or an electron is produced at the altitude of 10,000 meters and their energies are 10 GeV. It is well known that earth generates magnetic field. The strength of magnetic field is around 0.25 Gauss to 0.65 Gauss \cite{view1}. Charged particles with 10 GeV are bent by the magnetic field. Let us take the average value of earth's magnetic field of 0.45 Gauss (\(=4.5 \times 10^{-5} \) T). We calculate energy emission due to synchrotron radiation for muon and electron, respectively. \\
{ \bf (i) In case of a muon with 10 GeV } \\
A produced muon is bent by the earth magnetic field and its radius is calculated by the formula of (3),
% equation (6)
\begin{equation}
\rho =3.336\times \frac{10GeV}{4.5\times 10^{-5} T}=741\times 10^{5} [m] =741[km].
\end{equation}
Muon travels from 10 km height to the ground and its orbit length is calculated,
% equation (7)
\begin{equation}
\rho \theta =10000.3 [m],
\end{equation}
where $\theta$ is an orbit angle from 10 km to the ground. We can thus calculate the time of flight for muon from (7),
% equation (8)
\begin{equation}\\
T=\frac{\rho \theta }{c}=\frac{10000.3[m]}{2.9979\times 10^{8} [m/s]}=33.36 \mu s .
\end{equation}
We have mentioned that muon lifetime is 2.2 $\mu$s. However, muon with the energy of 10 GeV gets longer lifetime of 208$\mu$s, because of a relativistic particle. 
We input muon mass (=0.10566GeV) and another numerical values into (5),
% equation (9)
\begin{equation}
\frac{dP}{dt}=(2.8779 \times 10^{-1})\cdot \frac{1}{(mc^{2})^{4}}\cdot \frac{E^{4}}{\rho ^{2}}=(2.8779 \times 10^{-1})\cdot \frac{1}{(0.10566)^{4}}\cdot \frac{10^{4}}{(7.41\times 10^{5})^{2}}=4.2\times10^{-4}[eV/s].
\end{equation}
Muon radiates the radio emission during 33.36 $\mu$s and we obtain the total radio emission energy from (9),
% equation (10)
\begin{equation}
P=\frac{dP}{dt}\cdot T=(4.2\times 10^{-4})\cdot (33.36\times 10^{-6})=1.4\times 10^{-8} [eV].
\end{equation}
Thus obtained synchrotron radiation energy is very small for a muon.  \\
{ \bf (ii) In case of an electron with 10 GeV } \\
Let us estimate radiation length of electron at the altitude of 10 km. Particle Data Book \cite{Beringer} gives the value of radiation length at \( 20^{0}C \) on the ground: \(36.66 g/cm^{2} (=30423.2 cm) \). The atmospheric pressure reduces to 0.2677 atm at 10,000 meters height. We can obtain the radiation length of electron, provided the atmospheric pressure is assumed to be 0.2677 atm. We obtain the value of 
% equation (11)
\begin{equation}
L=\frac{30423.2cm}{0.2677atm}=113647cm=1136m.
\end{equation}
We may assume that an electron travels 1136 meters without any interactions. Even though electron with 10 GeV is bent by earth magnetic field, the curvature is negligible small. So the time of flight is obtained by
% equation (12)
\begin{equation}
T=\frac{1136m}{2.9979\times 10^{8} m/s}=3.79 \mu s.
\end{equation}
With the equation of (4), radio emission energy is given by
% equation (13)
\begin{equation}
\frac{dP}{dt}=(4.22\times 10^{12})\cdot \frac{E^{4}}{\rho ^{2}} [eV/s]=(4.22\times 10^{12})\cdot \frac{10^{4}}{(7.41\times 10^{5})^{2}}=7.686\times 10^{5} [eV/s].
\end{equation}
Total energy for radio emission is calculated by
% equation (14)
\begin{equation}
P=\frac{dP}{dt}\cdot T=(7.686\times 10^{5} eV/s)\cdot (3.79\times 10^{-6}s)=2.91 [eV].
\end{equation}
The result obtained in (14) is quite higher than the loss in (10). One can know that heavier charged particles than an electron emit radio emission due to synchrotron radiation, and the synchrotron radiation loss obtained above is negligible small comparing with that of electron. 

It concludes that charged particles radiate radio emission due to synchrotron radiation and electrons or positrons become main source for synchrotron radiation from air showers.

%% Subection 2.3
\subsection{ \bf Polarization of radio emission and coherent synchrotron radiation}
We have calculated the power of radio emission from charged particles in former sections. Let us consider the polarization of radio emission. The direction for the earth's magnetic field is defined from the South Pole to the North Pole. It must be considered that charged particles produced through strong and electromagnetic interactions in atmosphere have equally same numbers of positive and negative charges. Observed charge ratio regarding muons is given 1.27 \cite{Muraki}. It means that muons with positive charge excess a little. We neglect the charge ratio here, because electrons and positrons mainly generate radio emission. Since charged particles produced in higher altitude are bent by the earth magnetic field, particles with positive charge go toward east side and those with negative charge travel for west side. Charged particles consequently split into the east and west directions. Therefore radio emissions from air showers are polarized toward the east-west direction. In order to measure the polarization, as an example, let us assume that one uses a Yagi-Uda antenna (see Appendix C) to detect radio emission. Yagi-Uda antenna is well known to detect radio emission signal for television. Let us assume that one sets up a Yagi-Uda antenna on the ground. The direction of antenna faces the east. An air shower event with its zero zenith angle occurs at the east place away from the antenna location. If one sets upright the metal rods of Yagi-Uda antenna, one can expect higher signal level from air shower. On the other hand, the metal rods are set horizontal direction and obtained signal level becomes lower. Thus we can know the polarization of radio emission.

We may expect the coherent synchrotron radiation from air shower events. It means that electrons and positrons produced through cascade shower form a disk 
 and let us assume the thickness "L" as shown in Fig.1. 
If the wavelength "$\lambda$" of radio frequency produced by radio emission is longer than the length "L" of thickness, 
coherent synchrotron radiation takes place. The power from radio emission becomes higher than that of normal synchrotron radiation. If we assume the number N of produced electrons and positrons, the power of radio emission is proportional to the number of N for normal synchrotron radiation. On the other hand, the coherent synchrotron radiation gives the power of $ N^{2}$. Therefore the power from radio emission becomes very high. This is also mentioned in Ref.12. One can easily understand that coherent synchrotron radiation does not occur in higher frequency region.
%
% Figure 1 should be inserted here
%Fig_1
% pdf file is OK
\begin{figure}
\begin{center}
% \caption{\label{fig1}}
    \includegraphics[scale=0.4]{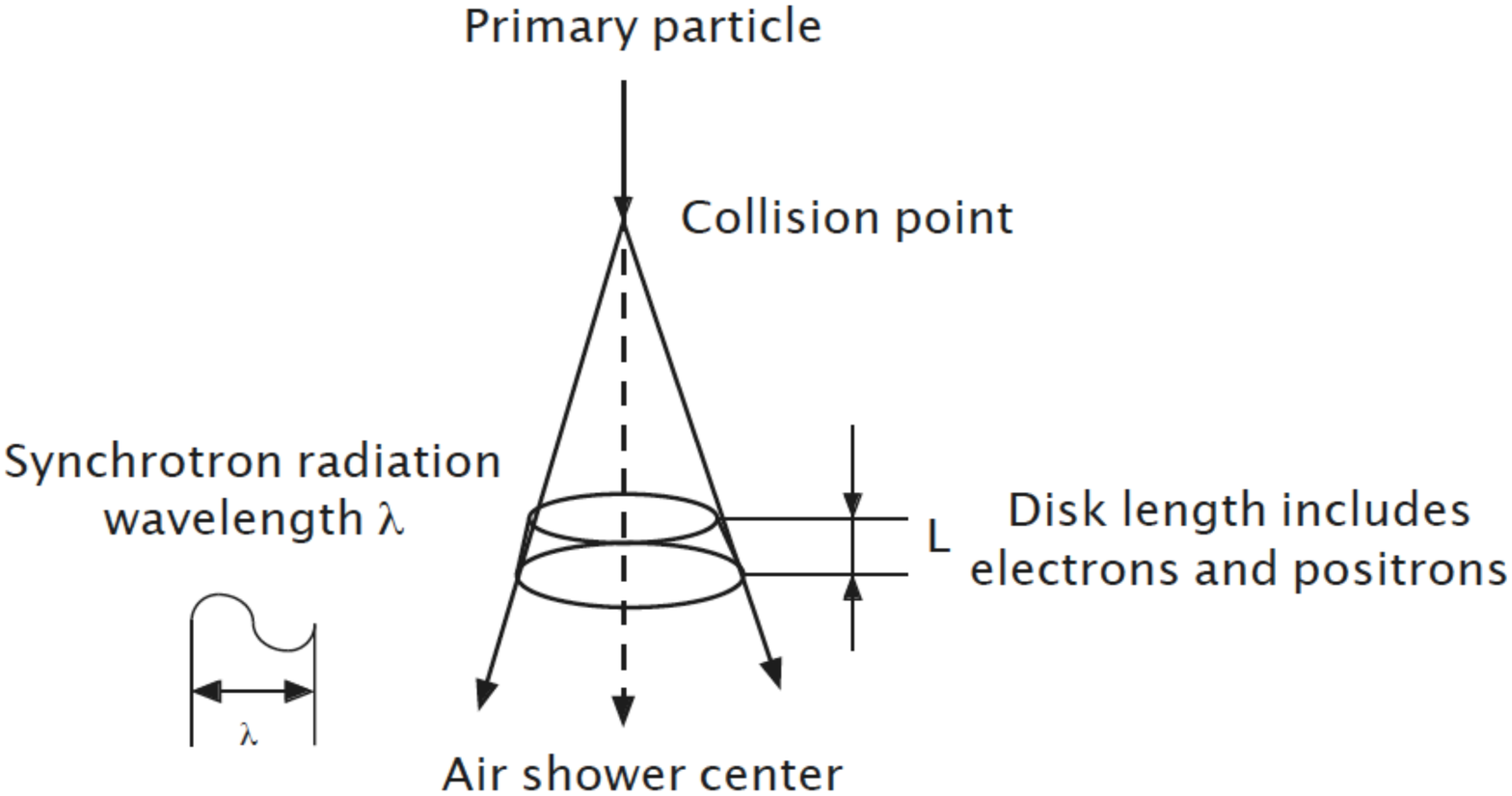} 
 \caption{ Coherent synchrotron radiation occurs when the wavelength "$\lambda$" of radio frequency is longer than the thickness "L" of disk, which contains
 many electrons and positrons.}
% \caption{Coherent synchrotron radiation occurs when the wavelength "$\lambda$" of radio frequency is longer than the thickness Ò L Ó of disk, which contains
% many electrons and positrons.}
\label{fig1}
\end{center}
\end{figure}

%% Subection 2.4
 %% Subection 2.4
\subsection{ \bf Frequency spectrum and electric field strength from radio emission}
Radio emission contains continuous radio frequency components. Fujii et al.  \cite{Fujii} showed the frequency spectrum from radio emission, however they did not calculate absolute energy intensity of radio emission. Huege et al.  \cite{Huege} have made a simulation and showed electromagnetic field intensity with respect to the radio frequency (RF) for the first time in the world. The spectrum obtained by Huege is useful to evaluate feasibility study for the detection of radio emission using antennas. Basing on their simulation results, we discuss the possibility for the detection of radio emission from air shower events. First of all, let us show the simulation result obtained by Huege. Using equation (2) in Ref.[12], electric field strengths with respect to radio frequency are shown in Fig.2, and also numerical values for electric field strengths from 200 MHz to 900 MHz are listed on Table 1. The air shower energy is \(1.0  \times 10^{17}  \) eV and data show electric field intensities for two points of 20 m and 140 m away from air shower center to north, respectively. Incident angle of a primary proton is zero which means vertical direction. In order to evaluate detectable threshold level, the numerical values as shown in Table 1 are useful later.

% Table 1
\begin{table}[ht]
\begin{center}
\begin{footnotesize}
%\begin{tiny}
\caption{Electric field strengths with respect to frequencies are listed \cite{Huege}. Air shower energy is \(1.0  \times 10^{17}  \) eV.}
%\begin{tabular}{llr}
%\begin{tabular}{lcc}
\begin{tabular}{|c| l| l| l| l| l| l| l| l|}
\hline
 Frequency in MHz  &  200  &  300  &  400  &  500  &  600  &  700  &  800  &  900    \\
 \hline
	20-m point  unit: $(\mu V/m)/(MHz)$  & 0.1682 & 1.958E-2 &  2.281E-3 & 2.655E-4 & 3.091E-5 & 3.599E-6 &  4.190E-7 & 4.879E-8  \\
 \hline
	140-m point unit: $(\mu V/ m)/(MHz)$  & 1.937E-3 & 3.206E-5 &  5.306E-7 & 8.781E-9 & 1.453E-10 & 2.405E-12 & 3.980E-14 & 6.586E-16 \\
\hline
\end{tabular}
\end{footnotesize}
%\end{tiny}
\end{center}
\end{table}
%% \end{center}
 
 %
% Figure 2 should be inserted here
%Fig_2
 \begin{figure}
\begin{center}
  \includegraphics[scale=0.5]{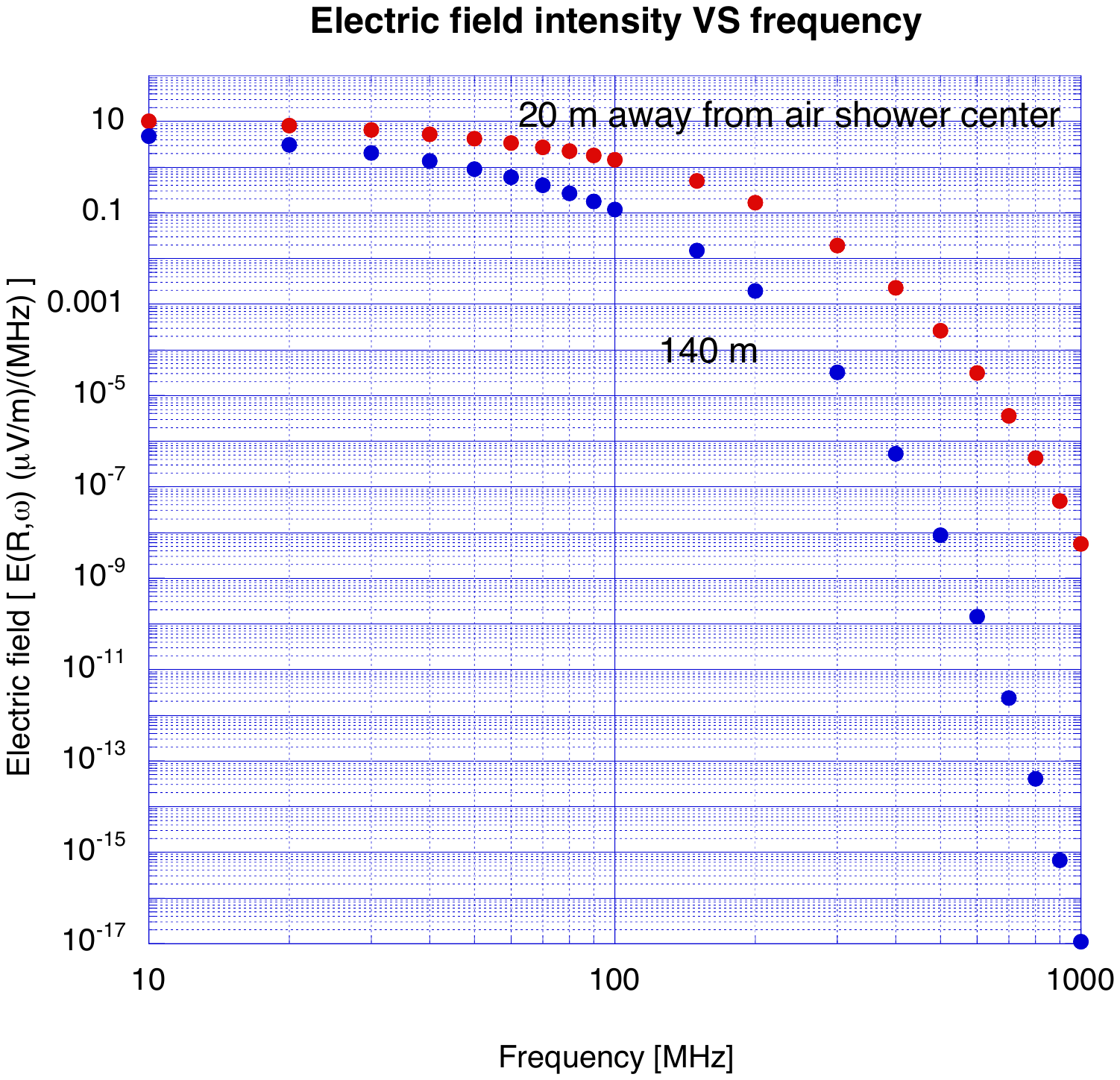}
  \caption{ Electric field strengths at the points of 20 m and 140 m away from the center of air shower \cite{Huege}. Air show energy is \( 1.0  \times 10^{17} \) eV.}
  \label{fig2}
  \end{center}
\end{figure}

%% Section 3
\section{Antenna to detect radio frequency leakage from accelerator facilities}
In order to evaluate the possibility of detection for radio emission from air shower events, it is useful to introduce an antenna, which is calibrated electric field strength in each frequency region. The purpose to use the calibrated antenna is to check the level of EMI (Electro-Magnetic Interference). The calibrated antenna is thus commercialized as a product. Therefore one can measure the absolute value of EMI. The calibrated antenna detects the leakage level of radio frequency (RF) from RF power transmitter system in accelerator facilities or other factories. The Radio Wave Management Law regarding RF or illegal RF generators stringently defines the limit of the leakage level and also prohibits the operation. Therefore, in order to suppress the level of RF leakage that is called Electro-Magnetic Compatibility (EMC), various calibrated antennas are available to measure the leakage level of RF power.
 
 If one purchases a commercialized calibrated antenna to detect RF leakage, one can find the specification that describes about available frequency region and antenna factor (see Appendix A). We have purchased a calibrated antenna HP11966D for the measurement of RF leakage in a synchrotron radiation facility in Japan. Let us show the relation between antenna factor and frequency in Fig.3. Since numerical values for the antenna factor are available for the evaluation of air shower, those values are read from Fig.3 and are listed numerical values in Table 2. The shape of antenna is a Yagi-Uda antenna (see Appendix C). One can know that the antenna covers the frequency region from 200 MHz to 1000 MHz as seen in Fig.3
 
 We completed the preparation to evaluate the possibility and threshold level for air shower detection using radio emission. In next section we explain the evaluation for the detection of air shower. 
 
 % Table 2
\begin{table}[ht]
\begin{center}
\begin{footnotesize}
\caption{Antenna factor vs frequency: HP11966D.}
%\begin{tabular}{llr}
%\begin{tabular}{lcc}
\begin{tabular}{|c| l| l| l| l| l| l| l| l|}
\hline
 Frequency in MHz  &  200  &  300  &  400  &  500  &  600  &  700  &  800  &  900    \\
 \hline
	Antenna factor  unit: ($dB\mu V/ m)/( dB\mu V)$  & 11.5 & 14.6 &  15.6 & 18.2 & 19.2 & 21.4 &  21.7 & 23.2  \\
 \hline
\end{tabular}
\end{footnotesize}
\end{center}
\end{table}
%% \end{center}

  %% Section 4
\section{Evaluation for air shower event detection using antennas}
 It must be more understandable to show concrete numerical values about electric field strength from air shower. Let us start the numerical treatment using Fig.3 or Table 2.
  First of all, we need to summarize the units about radio frequency in Appendix B. It is more convenient to use power unit dBm to handle RF. 
 
%% subsection 4.1
\subsection{\bf Detection of RF using a calibrated antenna: HP11966D}
At a synchrotron radiation facility, an antenna HP11966D has been used to check RF leakage from RF power transmitter systems. The concrete configuration for RF leakage measurement is shown in Fig.4. The calibrated antenna covers the frequency region from 200 MHz to 1 GHz as shown in Table 2. On the other hand, RF from air shower as shown in Fig.2 indicates that intensity of radio emission at lower frequency region is higher than that of higher frequency region. It means that the calibrated antenna HP11966D has disadvantage to detect higher frequency region. However it could be useful to evaluate the feasibility for the detection of radio emission using an antenna. Let us discuss quantitatively the feasibility.

A preamplifier with the gain of 25dB is installed. In order to analyze frequency spectrum and its amplitude for input signal, a spectrum analyzer is generally available. Let us summarize total attenuation value and gain for the preamplifier;
  \begin{equation}
Total \  \  attenuation \  \ of \   cable = \ (0.326 dB)+((0.600 dB)=0.926 dB,	
 \end{equation}
   \begin{equation}
   Total \  \   gain \  \  = 25 dB \  \  from \  \  a \  \  preamplifier.
 \end{equation} 
 Let us evaluate the detection of radio emission from air shower events step by step. \\
 % Step 1
{ \bf (1) Step 1} \\ 
 The unit to express electric field strength is presented by (see Fig.2 or Table 1),
   \begin{equation}
  ( \mu V/m )/(MHz).
 \end{equation} 
 It is not convenient to use this unit as mentioned above. One introduces more useful expression $(dB \mu V/m)$ as shown in Appendix B, (43). From Table 1, we convert the value at 200 MHz from 
 $(\mu V/m)$ to $(dB \mu V/m)$,
   \begin{equation}
  0.1682( \mu V/m)/(MHz) = 20 log(0.1682) = -15.48 (dB \mu V/m)/(MHz).  \\
 \end{equation} 
 % Step2
 { \bf (2) Step 2} \\ 
 The RF power at the input of the spectrum analyzer as shown in Fig.4 is calculated,
 % equation (15)
\begin{equation}
  Input \ power = -(antenna \ factor \ + \ cable \ attenuation) \ + \ (gain \ of \ preamplifier)
\end{equation}
   Let us input concrete values obtained in (15) and (16) at 200 MHz and antenna factor in Table 2.
 % equation (16)
\begin{equation}
   Input \ power = -((11.5) + (0.926 dB)) + 25 dB =12.57 dB \ at \ 200 MHz.
\end{equation}
The equation (20) does not include RF power detected by an antenna. In next step, we need to add RF power to (20).  \\
%Step 3
 { \bf (3) Step 3} \\
  In order to obtain the input RF power at the spectrum analyzer as shown in Fig.4, one needs to add power obtained in (18). It is more convenient to use RF power unit $ dBm $. To do that, one needs to add -107 dBm given in Appendix B, (49). Thus we obtain input power at the frequency of 200 MHz,
% equation (17)
\begin{equation}
Input \ power = [(12.574 dB) +(-15.48 dB \mu V/m) ] -107 dBm = -109.9 dBm/ MHz.
\end{equation}
Everyone wonders whether the value of -109.9 dBm/MHz is detectable or not. In order to answer this question, we have to know the natural background noise level. It is called thermal noise. Let us explain it in next section.

 In next section, we discuss about the threshold level to detect radio emission signal from air shower.

%
% Figure 3 should be inserted here
% Fig_3
 \begin{figure}
\begin{center}
  \includegraphics[scale=0.4]{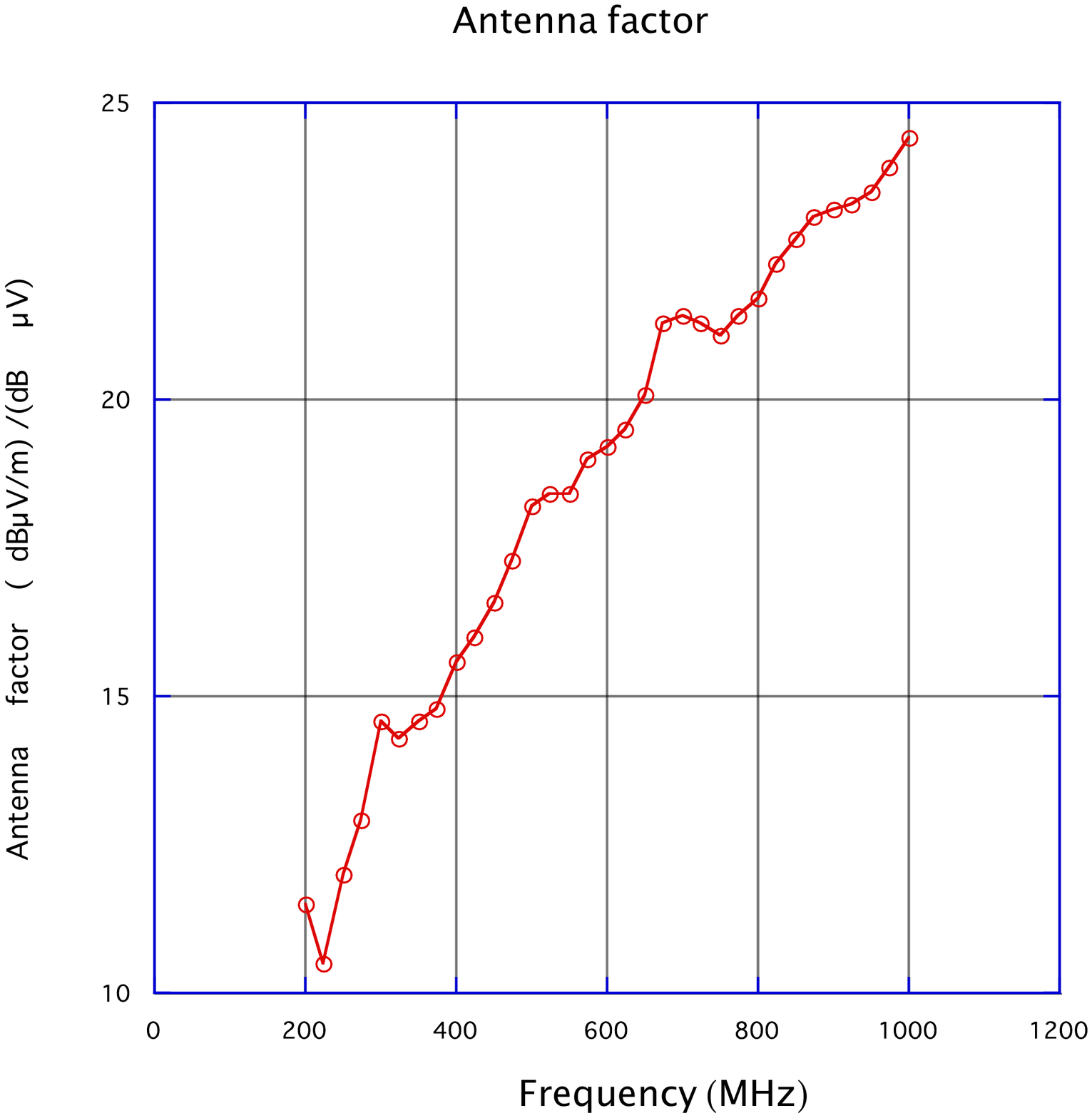}
  \caption{Relation between antenna factor and frequency for a calibrated antenna: HP11966D.}
  \label{fig3}
  \end{center}
\end{figure}

%
% Figure 4 should be inserted here
% Fig_4
\begin{figure}
\begin{center}
  \includegraphics[scale=0.4]{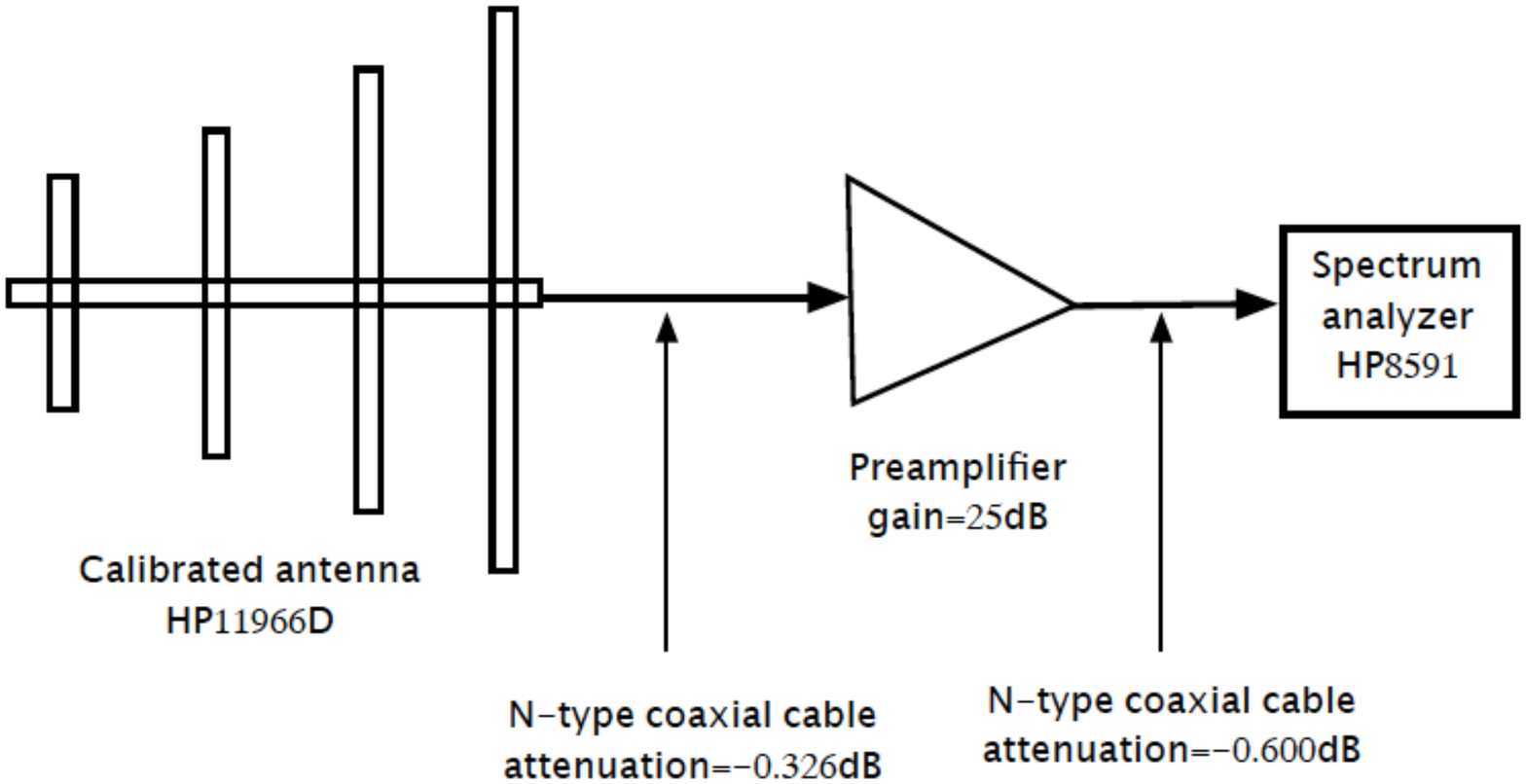}
  \caption{ Layout to measure RF leakage from RF power transmitter system at a synchrotron radiation facility is shown.}
  \label{fig4}
  \end{center}
\end{figure}
% 

%% subsection 4.2
\subsection{\bf Thermal noise}
If the RF power obtained by an antenna is as small as the thermal noise, we cannot detect radio emission from air shower. Therefore it may say that thermal noise (one can see any textbooks ) gives the threshold level for the detection of air shower events. The average voltage V due to the thermal noise is expressed by
% equation (18)
\begin{equation}\
<V^{2}>=4k\cdot T\cdot R\cdot \Delta f
\end{equation}
\hspace{3cm} k :  Boltzmann constant ( \( 1.38\times 10^{-23}JK^{-1}\) ),\\
\hspace*{3cm} T : T: room temperature ( ~300 K ),\\
\hspace*{3cm} R : R: input impedance ( 50 $\Omega $) ,\\
\hspace*{3cm} $\Delta f$ :  frequency width ( Hz).\\ 
Let us discuss the relation between thermal noise and simulation result as shown in Fig.2 and Table 1. In the equation of (22), $\Delta f$ is unknown parameter. How can we know that? The clue is found in Table 1 and Fig.2. The strength of electric field is expressed in the unit of $ (\mu V/m)/(MHz) $. It means that bandwidth is 1 MHz. We can set the $ \Delta f =1MHz=1.0  \times 10^{6} Hz  $. Thus we calculate the equation (22) at 300 K,
% equation (19)
\begin{equation}
<V^{2}>=4\times (1.38\times 10^{-23})\times 300\times 50\times (1.0\times 10^{6})=8.28\times 10^{-13}.
\end{equation}
To obtain the power from (23), we can use the formula (46) in Appendix B. The power of thermal noise is given by
% equation (20)
\begin{equation}
P=\frac{V^{2}}{R}=\frac{8.28\times 10^{-13}}{50}=1.656\times 10^{-14}W=1.656\times 10^{-11}mW.
\end{equation}
We convert (24) to dBm unit by using formula (39) in Appendix B,
% equation (21)
\begin{equation}
P=10 \times log(1.656\times 10^{-11})=-107.8dBm.
\end{equation}
Since the preamplifier as shown in Fig.4 amplifies the thermal noise, we have to add the gain of preamplifier to (25) and finally obtain thermal noise level at the input of the spectrum analyzer as shown in Fig.4.
% equation (22)
\begin{equation}
P=-107.8dBm+25.0dBm=-82.8dBm.
\end{equation}
We can compare the data of (21) with (26). 
Let us summarize as following; \\
{ \bf  (I)}  If one tries to detect the radio emission at 200 MHz obtained by Huege  \cite{Huege}, using the antenna as shown in Fig.4, the RF power level is obtained at the input of a spectrum analyzer,
% equation (23)
\begin{equation}
Obtained \ \ RF \ \ power \ \ level = -109.9 dBm \approx -110 dBm/MHz \ \ at \ \ 200 MHz,
% Obtained \ \ RF \ \ power \ \ level = -109.9 dBm \ \ at \ \ 200 MHz,
\end{equation}
where primary cosmic ray energy is $ 1.0 \times 10^{17} eV$ . \\
 { \bf (II)}  Thermal noise level is given by 
% equation (24)
\begin{equation}
Thermal \ noise \ at \ 300 K= -82.8 dBm. 
\end{equation}
 This value of -82.8 dBm is observable. In fact, we directly observed the noise level using the detection system as shown in Fig.4. 

As one sees the results of (27) and (28), one can understand that it is impossible to detect air shower signal at 200 MHz. It means that the detection of radio emission at higher frequency region more than 200 MHz is quite difficult as shown in Fig.2. 

We summarize calculated results in higher frequency region in Table 3. Detection of air shower signals gets more difficult as the distance becomes far away from the center of air shower events.

% Table 3
\begin{table}[ht]
\begin{center}
\begin{footnotesize}
%\begin{tiny}
\caption{RF power at various frequency regions in detecting radio emission with a calibrated antenna as shown in Fig.4. Primary cosmic ray energy is $ 1.0 \times 10^{17} eV $.}
%\begin{tabular}{llr}
%\begin{tabular}{lcc}
\begin{tabular}{|c| l| l| l| l| l| l| l| l|}
\hline
 Frequency in MHz  &  200  &  300  &  400  &  500  &  600  &  700  &  800  &  900    \\
 \hline
	Point at 20 m unit: ($ \mu V/ m)/( MHz)$  & 0.1.682 & 1.958E-2 &  2.281E-3 & 2.655E-4 & 3.091E-5 & 3.599E-6 &  4.190E-7 & 4.879E-8  \\
 \hline
        Point at 20 m unit: ($ dB \mu V/ m) $ in (18) & -15.48 & -34.16 &  -52.84 & -71.52 & -90.20 & -108.9 &  -127.6 & -146.2  \\
 \hline
 Input power at spectrum ananyzer   & \  & \  &  \  & \  & \ & \  &  \  & \   \\
  unit : dBm in (19) and (20)   & -110 & -132 &  -151 & -173 & -192 & -213 &  -232 & -255  \\
  \hline
\end{tabular}
\end{footnotesize}
%\end{tiny}
\end{center}
\end{table}
%% \end{center}

%% subsection 4.3
\subsection{\bf Effect of band-pass filter to detect radio emission from air shower events}
In former section, we obtained a negative result. It is useful to consider the detection of radio emission furthermore. One may simply say that RF power level at 200 MHz point is too small to detect. In order to detect radio emission from air shower events, we need to know the relation between bandwidth and thermal noise. In order to detect radio emission from air showers, the former researchers as mentioned in section 1 did not consider about the bandwidth of a band-pass filter. However, to identify the air shower events from natural or artificial noise, we need to discuss about band-pass filter. The detail on the identification of air shower events is discussed in section 8. In order to know the relation between the bandwidth and the RF power level of radio emission, let us consider the relation between the bandwidth of detector and thermal noise.

We assume that the antenna as shown in Fig.4 accepts wider frequency region and the antenna factors in any frequency points are equal. As we mentioned, thermal noise level varies depending on bandwidth as seen in the equation of (22). We calculated RF power level at 200 MHz and 20 m point away from the center of air shower as shown in Table 1. The electric field strength is given
 by 0.1682 ( $\mu V/m $)/(MHz). Let us install an imaginary band-pass filter after the preamplifier in Fig.4. When we make the bandwidth wider and wider, the detected RF power level becomes higher depending on the bandwidth of band-pass filter. The calculated results are plotted in Fig.5. The result implies very important information to detect air shower events. If an antenna with wider bandwidth is available, air shower events might be detectable. In particular, RF power from air shower events in lower frequency region than 100 MHz becomes higher as shown in Fig.2.
To check the RF power level at 140 m point away from the center of air shower events, let us calculate the power level by changing bandwidth. The power level at 200 MHz is given by 1.973E-3 ( $\mu V/m $)/(MHz) from Table 1 and results are also shown in Fig.5.

It could be useful to discuss the bandwidth for a band-pass filter. A narrower band-pass filter suppresses thermal noise. However if one installs a narrower band-pass filter with $80 MHz \pm 1 MHz$, as an example, what happens? In order to obtain the answer, we have to know the time duration of air shower events. Electrons and positrons mainly generate the power of radio emission as explained in section 2.2 and they fly at the light speed. Therefore we may imagine that electrons and positrons are moving with electro-magnetic field of radio emission. The time duration of radio emission could be within $1 \mu s$. The simulation done by Huege et al. \cite{Huege}  shows that the time duration for radio emission is around 100 ns. It is very short time. Under such short time duration of air shower signal, the narrower band-pass filter would filter out real signal from air shower events. Since the bandwidth for a band-pass filter has a physical meaning, we have to carefully select the bandwidth. 

In order to know the effect of bandwidth for RF detector, we have to know the character of various antennas. In next section, we describe various antennas. 

 %
% Figure 5 should be inserted here
%  Fig_5
\begin{figure}
\begin{center}
  \includegraphics[scale=0.5]{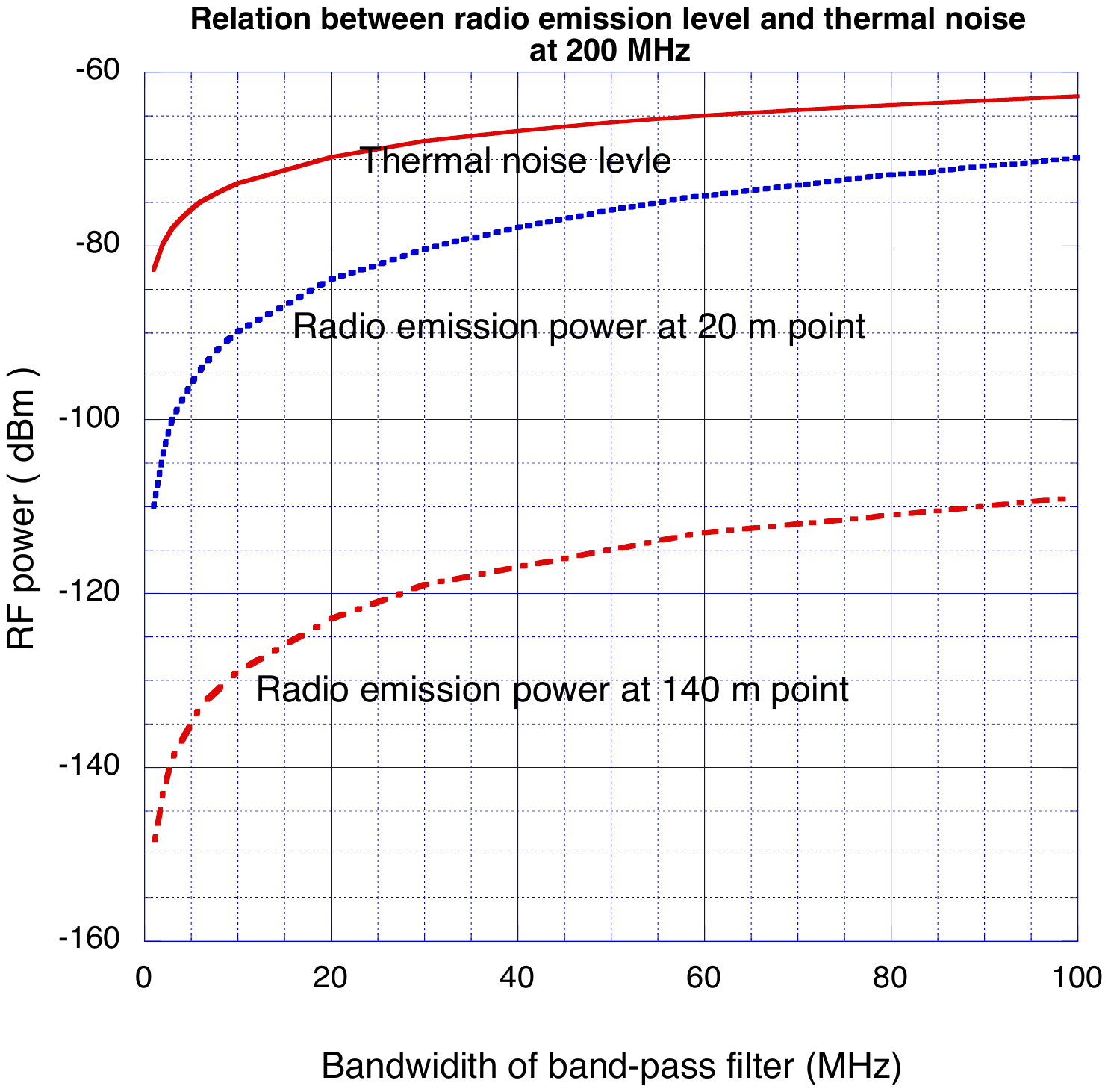}
  \caption{ Relation between thermal noise and bandwidth of a band-pass filter is indicated. The bandwidth becomes wider and wider, and RF power level at 200 MHz as shown in Fig.2 becomes higher. The data at 20 m and 140 m points are shown. Primary cosmic ray energy is  \( 1.0  \times 10^{17} \) eV. }
  \label{fig5}
  \end{center}
\end{figure}

%% Section 5
\section{Various antennas for the detection of radio emission}
In order to detect radio emission from air shower events, one example has been shown in former section. There are various antennas to detect radio emission from air shower. An antenna as shown in Fig.4 is one of the examples. The purpose of antenna is to detect RF leakage (EMC) and it should cover wide frequency range from 200 MHz to 1000 MHz. Thus, the detection efficiency for each frequency is not good. As one sees the structure of antenna as shown in Fig.4, the rod lengths are different. The different rods with different lengths correspond to different resonance frequencies. In order to detect a specified frequency effectively, all lengths of rods should have same one. According to a textbook \cite{Kraus} , the best antenna is a parabolic dish antenna, which is expensive. Next candidate is an axial-mode helix antenna. Yagi-Uda and rectangular horn antennas have lower detection efficiency. Since the Yagi-Uda antenna is inexpensive, one can see the Yagi-Uda antennas on the roof of houses to detect RF for television. Those four antennas \cite{Kraus} are summarized in Appendix C. 

The most important point to detect RF is an antenna gain. The antenna gain has a relation with directivity D. The directivity is defined by  \cite{Kraus},
% equation # (25)
\begin{equation}
D=\frac{P(\Theta ,\Phi )_{MAX}}{P(\Theta ,\Phi )_{AVERAGE}}=\frac{4\pi }{\Omega _{A}},
\end{equation}
\hspace{3cm} $ \Omega _{A}$ :  beam area, solid angle. \\
Using this directivity, the gain G is represented,
% equation # (26)
\begin{equation}
G=k\cdot D ,
\end{equation}
\hspace{3cm} k : efficiency factor, $( 0\leq k\leq 1 ). $ \\
The gain is proportional to the directivity D. From Fig.16 in Appendx C, one can know that parabolic disk antenna has the best gain. We compare Yagi-Uda with axial-mode helix and know that the axial-mode helix antenna has twice larger gain than that of Yagi-Uda antenna. The rectangular horn antenna has a smaller directivity than those of parabolic disk and axial-mode helix antennas.

In order to detect air shower events, the antenna gain is one of the most important factors. However, air shower events randomly occur and the incident directions are not fixed. One can know that the better directivity D has a narrower solid angle $\Omega_{A}$ from the equation (29). It means that antennas with better directivity accept RF signal from narrower solid angle. Thus antennas with higher directivity are not always better detector for air shower events. On the other hand, one can also know bandwidth from Fig.16 in Appendix C. Yagi-Uda has a 15 $\%$ bandwidth. On the other hand, axial-mode helix has wider bandwidth of 71 $\%$. To capture the signals from air shower events, the wider bandwidth is more useful as mentioned in section 4.3 and attractive as a detector of air shower events. Thus it is not easy way to select antenna to detect RF from air shower events.

To understand actual sizes of antennas, we calculated them basing on Fig.16 in Appendix C and show their lists for antenna parameters in Table 4 for axial-mode helix antenna and Table 5 for Yagi-Uda antenna. The frequency region is selected from 10 MHz to 100 MHz based upon the spectrum as shown in Fig.2.

In order to detect RF from air shower, one can know the antenna size from Table 4 and Table 5. As an example, to detect the RF with 10 MHz, the antenna lengths "L" for both antennas are almost 45 meters long. It is not realistic size. On the other hand, let us take a look of 100 MHz frequency point. Both antenna lengths are around L=4.5 meters long. It seems to be reasonable size. If one sets the center frequency of antenna at 80 MHz, the axial-mode helix antenna covers wide frequency region of 71 $\%$, in which detectable power decreases by 3 dB. On the other hand, Yagi-Uda antenna has 15 $\%$ bandwidth. In order to compare bandwidths for both antennas, their bandwidths with the center frequency of 80 MHz are shown in Fig.6.

Yagi-Uda antenna has narrower bandwidth and antenna itself has a function of band-pass filter. On the other hand, axial-mode helix antenna has wider bandwidth. In order to detect radio emission effectively, an antenna with wider bandwidth has a merit and we need to describe RF detection method in next section.

% Table 4
%% \begin{center}
%% \begin{table}[tb]
\begin{table}[ht]
\begin{center}
\begin{footnotesize}
\caption{Parameters for axial-mode helix antennas to detect frequency region from 10 MHz to 100 MHz. Refer to Fig.16 in Appendix C about definitions for various parameters. $\lambda$ stands for one wavelength.}
%\begin{tabular}{llr}
%\begin{tabular}{lcc}
\begin{tabular}{|c| l| l| l| l| l| l| l|}
\hline
 f  (MHz)  &  $\lambda (m) $  &   $\lambda/4 (m) $ &   $\lambda /\pi (m)$  &   $3\lambda /8 (m)$  &   $3\lambda/4 (m)$  &  L (m)  &  D      \\
 \hline
  10           & 29.98   &      7.495     &   9.543     &   11.24    &    22.49    &   44.97     &  18      \\
 \hline
  20	        &  14.99  &      3.748	  &    4.771     &   5.621    &    11.24     &   22.49	 &   18     \\
 \hline
 30	        &  9.993   &     2.498     &    3.181     &   3.747    &     7.495    &    14.99	  &  18       \\
 \hline
  40	         &  7.495   &    1.874     &    2.386      &   2.811     &    5.621   &     11.24    &   18     \\
 \hline
   50	         &   5.996   &    1.499     &	1.909	       &   2.249    &    4.497    &	8.994	   &   18     \\
 \hline  
  60	         &   4.997   &     1.249     &  1.591       &    1.874    &    3.748    &    7.496   &   18     \\
 \hline  
 70	         &   4.283   &     1.071     &   1.363     &    1.606     &    3.212    &    6.425    &  18    \\
 \hline
 80	         &   3.747   &     0.9368    &   1.193     &   1.405    &    2.810     &    5.621    &   18    \\
  \hline
  90	         &   3.331   &    0.8328     &    1.060    &    1.249    &   2.498     &   4.997     &   18  \\
  \hline
    100	&   2.998   &    0.7495     &    0.9543   &   1.124    &    2.249     &  4.497     &   18  \\
  \hline
\end{tabular}
\end{footnotesize}
\end{center}
\end{table}
%% \end{center}

% Table 5
%% \begin{center}
%% \begin{table}[tb]
\begin{table}[ht]
\begin{center}
\begin{footnotesize}
\caption{Parameters for Yagi-Uda antenna are calculated from 10 MHz to 100 MHz. Definitions about all parameters are shown in Fig.16 in Appendix C.}
%\begin{tabular}{llr}
%\begin{tabular}{lcc}
\begin{tabular}{|c| l| l| l| l| l| l| l| l| l| l| l| l|}
\hline
 f  (MHz)  &  $\lambda (m) $  &   L1 (m)  &   L2 (m)  &   L3 (m)  &   L4 (m)  &  L5 (m)  &  L6 (m)   &  $\lambda/100 (m) $   & $\lambda/4 (m)$   &  $\lambda/\pi (m)$  &  L (m) & D      \\
 \hline
  10	        &  29.98      &   14.39   &   13.79    &   13.19     &  13.19   &  12.89   &   11.99   &   0.2998   &  7.495    &  9.543    & 45.67	    &   9.1    \\
 \hline
  20	        &  14.99     &   7.195   &  6.895   &   6.596   &   6.596   &  6.446   &  5.996   &   0.1499   &   3.748   &    4.771   &   22.83   &   9.1     \\
 \hline
 30	        &   9.993    &   4.797    &  4.597   &  4.397    &   4.397   &   4.297   &  3.997  &   9.993E-2  &  2.498  &  3.181    &   15.22   &   9.1   \\
 \hline
 40	        &   7.495    &   3.598     &  3.448   &   3.298   &   3.298   &  3.223    &   2.998  &   7.495E-2   &   1.874   &   2.386  &   11.42  &  9.1     \\
 \hline
   50	        &   5.996    &   2.878      &  2.758   &   2.638   &   2.638   &  2.578   &  2.398   &  5.996E-2     &   1.499    &  1.908  &  9.133   &    9.1   \\
 \hline  
  60	        &  4.997     &   2.399      &  2.298    &  2.198    &  2.198    &    2.149  &  1.999	 &  4.997E-2    &   1.249    &  1.591    &   7.611  &   9.1     \\
 \hline  
 70	        &   4.283    &  2.056      &   1.970    &   1.885   &  1.885     & 1.842     &  1.713   &  4.283E-2   &    1.071   &  1.363   &  6.524    &  9.1    \\
 \hline
  80	         &   3.747    &  1.799     &   1.724     & 1.649     &  1.469    &  1.611    &  1.499    &   3.747E-2    &  0.9368    &  1.193    &  5.708    &  9.1    \\
  \hline
  90	          &  3.331    &  1.599    &  1.532   &   1.466   &  1.466    &  1.432   &   1.332   &   3.331E-2   &   0.8327   &  1.060    &  5.074         &  9.1  \\
  \hline
   100	          &   2.998    &  1.439    &  1.379   &  1.319   &   1.319   &  1.289   &   1.199   &  2.998E-2   &   0.7495   &  0.9543    &  4.567   &  9.1 \\
  \hline
\end{tabular}
\end{footnotesize}
\end{center}
\end{table}
%% \end{center}

 %
% Figure 6 should be inserted here
%  Fig_6
\begin{figure}
\begin{center}
  \includegraphics[scale=0.5]{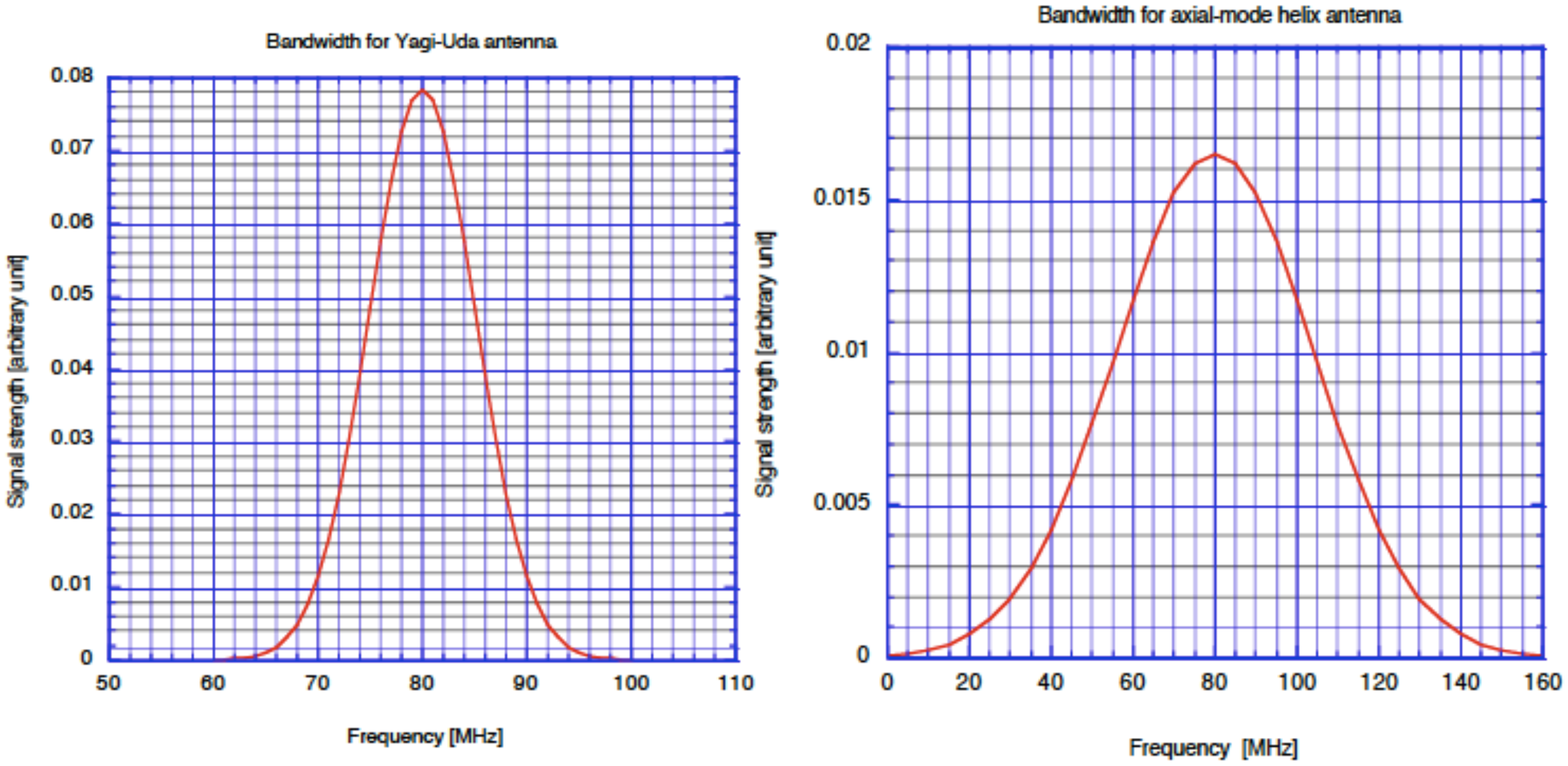}
  \caption{Bandwidths for Yagi-Uda antenna (left figure) and axial-mode helix antenna (right figure) with the center frequency of 80 MHz are shown. The bandwidth for Yagi-Uda antenna is narrower than that of axial-mode helix antenna. To obtain higher signal from air shower events, the wider bandwidth is the better.}
  \label{fig6}
  \end{center}
\end{figure}

%% Section 6
\section{RF detection method}
In this section, we describe RF detection method. An antenna detects radio emission and the received signal should be converted from high frequency of alternating current (AC) to direct current (DC) level. RF detectors are usually used to convert AC to DC levels. To explain RF detection method, one example is shown in Fig.7.

 %
% Figure 7 should be inserted here
% Fig_7
\begin{figure}
\begin{center}
  \includegraphics[scale=0.5]{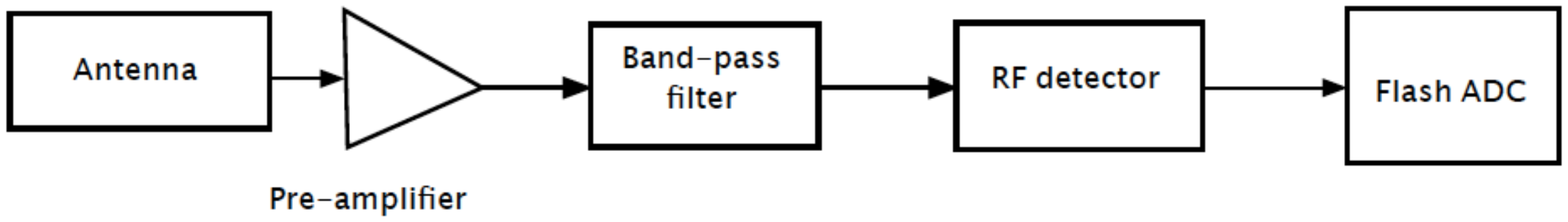}
  \caption{ Fundamental RF detection method is shown. RF detector is placed in front of a flash ADC (Analog to Digital Converter). To suppress thermal noise, a band-pass filter is installed. }
  \label{fig7}
  \end{center}
\end{figure}

The output from the antenna is connected to a preamplifier. In order to suppress thermal noise as discussed in section 4.3, a band-pass filter is installed behind the preamplifier. The band-pass filter should have the same center of frequency as the antenna and also should cover enough bandwidth. If one adopts Yagi-Uda or axial-mode helix antenna, the band-pass filter should completely cover the frequency region of an antenna.

RF is converted to DC level and furthermore the DC voltage is converted from analog to digital signal as seen in Fig.7. One can see the name of flash ADC, which converts analog signal to digital one in a short time. As an example, if the operating clock for the flash ADC is 100 MHz, the analog signal is converted into digital signal every 10 ns. The reason to select the flash ADC is to measure the incident direction of air shower events as well as the energy of air shower event at the same time. If one uses several antennas, one can know the time differences between antennas. The arrival timing gives the incident direction of air shower events. 

If an antenna covers a wide frequency region like an axial-mode helix antenna, we can set a few different band-pass filters as shown in Fig.8. The reason to adopt several band-pass filters is related to a method for the identification of air shower events from noises. It is discussed in section 8. 

We detect RF powers at different frequency points at the same time. As an example, one sets up the center frequency at 80 MHz for an axial-mode helix antenna. The band-pass filter is selected $80 MHz \pm 5 MHz $ and another two filters are respectively selected like $60 MHz \pm 5 MHz $, $100 MHz \pm 5 MHz $. As one sees bandwidth for an axial-mode helix antenna in Fig.6, detection efficiency at both sides of 60 MHz and 100 MHz drops by around 29 $\%$ comparing with that of center frequency of 80 MHz. In this model, only one antenna is manufactured and three different circuits should be prepared.

If three independent antennas with different frequency sensitivities are available, we can design another system as shown in Fig.9. This configuration has two merits. Firstly there is no power divider as shown in 
Fig.8, where RF power from a preamplifier is divided into three. So we can estimate the power loss of 6 dB due to power divider. Secondly the other merit is that we do not need to make a correction for antenna bandwidth. Three independent RF powers detected with three different antennas would represent the same spectrum as shown in Fig.2. Thus if the construction fund is good enough, independent detection system must be better.

 %
% Figure 8 should be inserted here
% Fig_8
\begin{figure}
\begin{center}
  \includegraphics[scale=0.5]{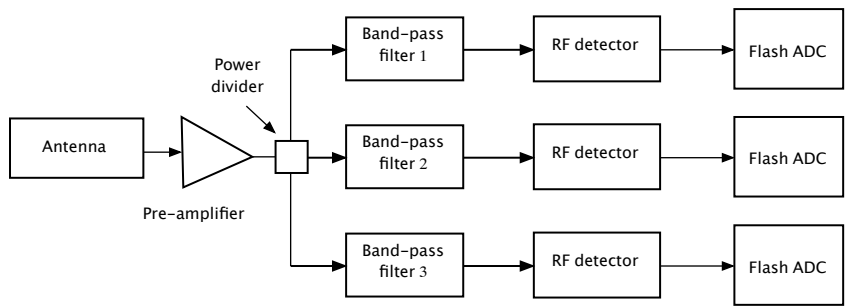}
  \caption{This block diagram shows that signal from an antenna is divided into three and each channel has a different band-pass filter to detect different frequency. }
  \label{fig8}
  \end{center}
\end{figure}

 %
% Figure 9 should be inserted here
% Fig_9
\begin{figure}
\begin{center}
  \includegraphics[scale=0.5]{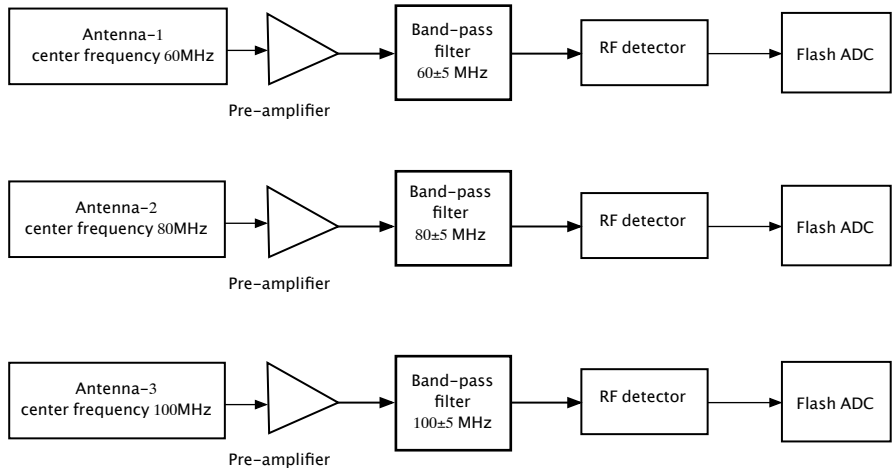}
  \caption{Three different antennas with different center frequencies, 60 MHz, 80 MHz and 100 MHz are set up. This configuration is easier to handle, since power attenuation is only cable and band-pass filter. }
  \label{fig9}
  \end{center}
\end{figure}

To convert from RF to DC voltage, RF detector is commonly used. We show an example for RF detector: MAX2016. The product accepts from low frequency to 2.5 GHz. In order to know the dynamic range, we have tested the RF detector at 508.58 MHz. The obtained dynamic range is shown in Fig.10. The RF detector covers from -75 dBm to 0 dBm in Fig.10. However the specification for MAX2016 says that the acceptable maximum input power is 10 dBm. Thus the actual dynamic range is over 80 dBm. RF is converted to DC voltage and is analyzed by a flash ADC (=Analog to digital converter). The flash ADC should be operated at the rate of 200 MHz or higher than 200 MHz clock, since the time duration of cascade shower would be expected around 100 ns. The flash ADC might analyzes the development process of air shower.
 %
% Figure 10 should be inserted here
% Fig_10
\begin{figure}
\begin{center}
  \includegraphics[scale=0.5]{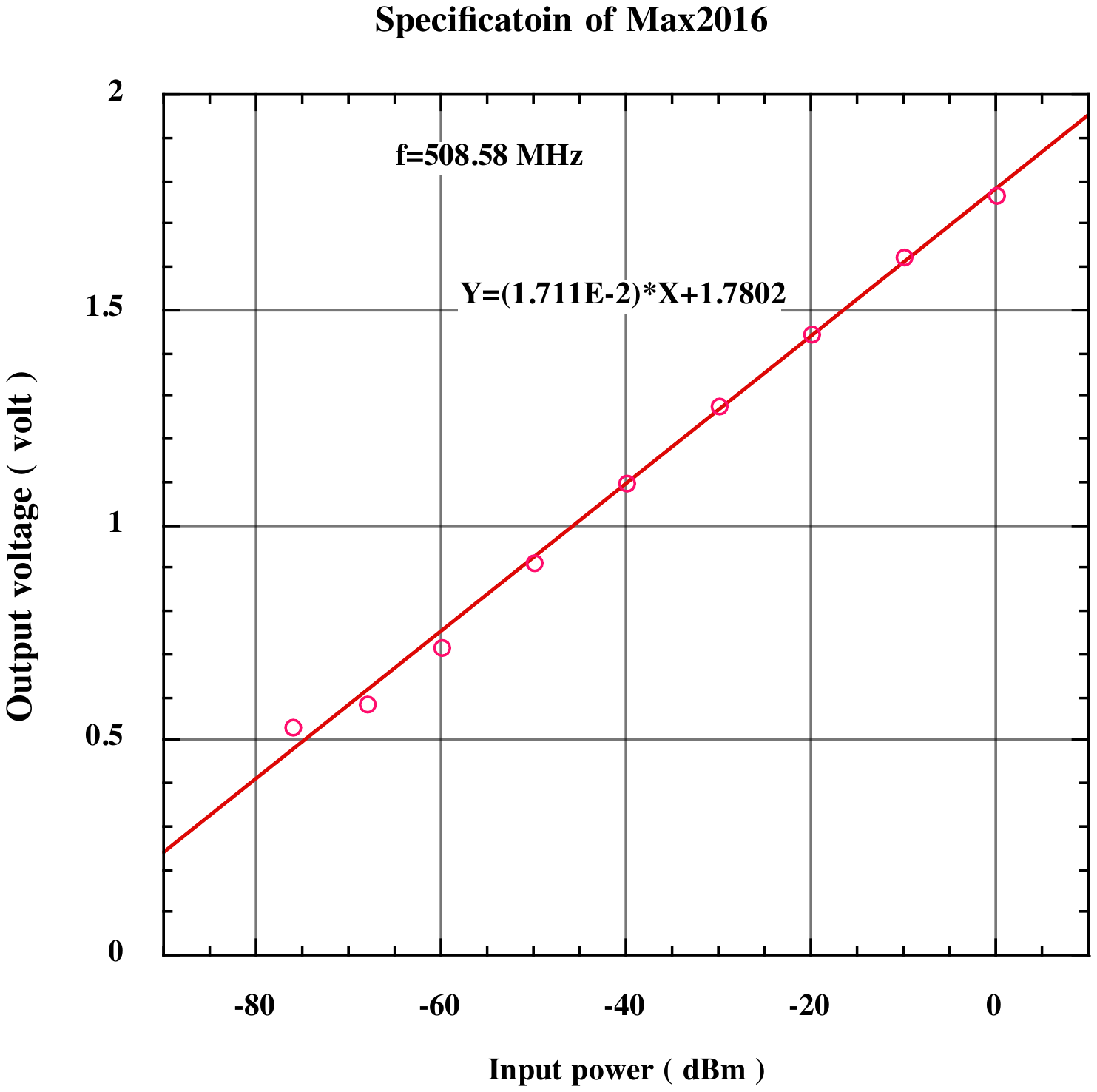}
  \caption{Dynamic range for an RF detector of MAX2016 was tested in a laboratory at 508.58 MHz. }
  \label{fig10}
  \end{center}
\end{figure}

%% Section 7
\section{Self-trigger method for radio emission}
So far, the detection of radio emission from air shower events has been generally supplied a trigger signal from scintillation array system. As a result, radio emissions from air shower have been detected. Let us mention about self-trigger method for radio emission. The total system basing on Fig.9 is shown in Fig.11. Analog output voltage from an RF detector is divided into two. One of the analog signals is inputted into a discriminator and the other is driven into a flash ADC. We set the threshold level in three discriminators. The threshold level should be set in accordance with the RF power level of air shower events, which should be calibrated in a laboratory as shown in Fig.10. The output signals from discriminators are connected to an AND-logic circuit. We can select any two or three folds coincidence. If the logic meets the set-up condition, the AND-logic circuit generates a signal. The output digital signal from the AND-logic circuit is delayed more than 100 ns by a gate delay generator, because the time duration of air shower events is expected around 100 ns \cite{Huege}. The delayed signal triggers three flash ADCs. Those flash ADCs are always converting input analog signal to digital data and are storing the data in the memory at the rate of a given clock. Once flash ADC accepts the trigger signal, and conversion process is stopped by the trigger signal from the AND-logic circuit. We obtain the past data with the time duration of $\mu s$ or longer before accepting trigger signal. Thus we can create trigger signal inside of the air shower detection system. In order to suppress accidental coincidence trigger signals, it is better to take two or more folds coincidence.

If flash ADCs are not used for the signal detection, we need to insert analog delay lines between RF detector and a simple ADC (there are two different type ADCs: a charged ADC, a peak detection ADC). Thus total system with flash ADCs becomes simpler. As the other merit adopting the flash ADC, if we set several or more antenna detection systems on the ground, we can measure the incident direction of air shower events from the information of time difference between different antennas. Since the flash ADC continuously stores data on memory, we simultaneously obtain both signal intensity and time stamp when air show events occur. The clock adopted by the flash ADC defines the time accuracy. If one wants more accurate incident direction of air shower events, we should set higher frequency clock for the flash ADC.
 %
% Figure 11 should be inserted here
% Fig_11
\begin{figure}
\begin{center}
  \includegraphics[scale=0.5]{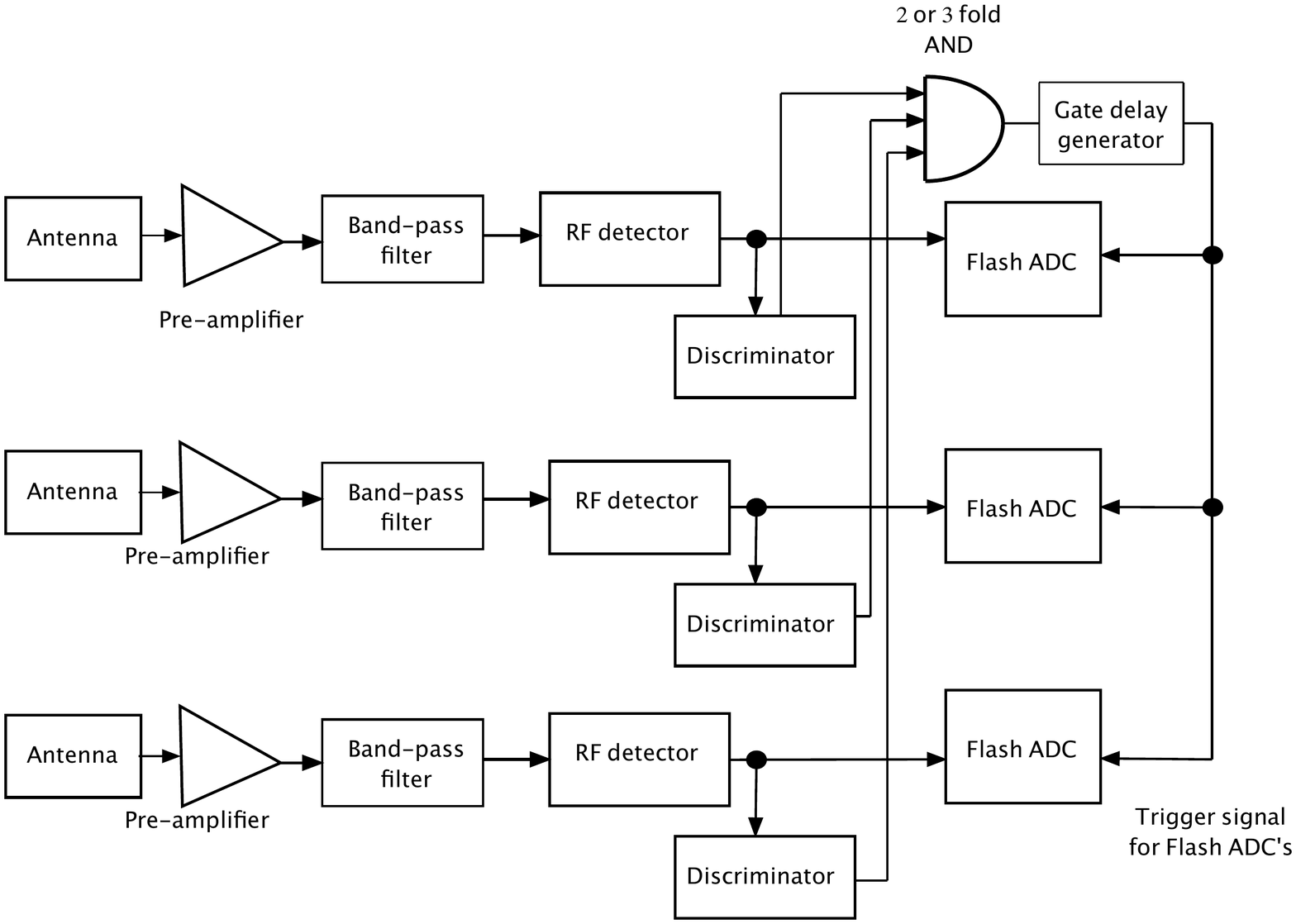}
  \caption{A block diagram to generate a self-trigger signal is shown. The flash ADCs continuously convert analog signal to digital signal and store digital data into memory. Flash ADCs detect a trigger signal and stop the conversion. In this system, one can measure both RF power level and time differences between antenna systems at the same time. }
  \label{fig11}
  \end{center}
\end{figure}

%% Section 8
\section{Identification of air shower events}
In order to identify radio emission of air shower events from natural or artificial noises such as lightning and meteor, Huege \cite{Huege} described the method, which is to measure the polarization of radio emission. As discussed in section 2.3, charged particles are bent to the directions of east and west by the earth's magnetic field. The electric field strength of radio emission on the polarized plane becomes larger than that on the plain with a right angle. On the other hand, we described another method in section 7, in which an antenna or several antennas detects RF powers at three or more different frequency points. Radio emission from air shower has a specific frequency spectrum as shown in Fig.2. If we could simultaneously detect radio emission at various different frequency points, we would have the same kind of frequency spectrum like Fig.2. Therefore antenna system that we showed in Fig.8 or Fig.9 is designed to detect radio emissions at different frequency points.

Let us mention the merit to detect radio emission in different frequency points. We assume the RF detection system as shown in Fig.11. We also assume that one selects three different axial-mode helix antennas, whose center frequencies are 60 MHz, 80 MHz and 100 MHz, respectively. Under the condition, let us assume that antenna system detects air shower events or something noise like lightning. If the detected event has the same spectrum as shown in Fig.2, we might say that it could be true radio emission from air shower events. On the other hand, since lightning event is a kind of arc in the atmosphere, obtained frequency spectrum is expected to have a peak frequency. If the time period for lightning is given by $\Delta$t, the center of frequency f is calculated as f=1/$\Delta$t. In order to obtain the frequency spectrum, we carry out Fourier transform. Let us assume that the amplitude of lightning has the unit of one and the time structure has a rectangular form as shown in Fig.12. We can carry out the calculation. 
% It should notice that the suffix "i" stands for imaginary part. 
%
% Figure 12 should be inserted here
% Fig_12
\begin{figure}
\begin{center}
  \includegraphics[scale=0.3]{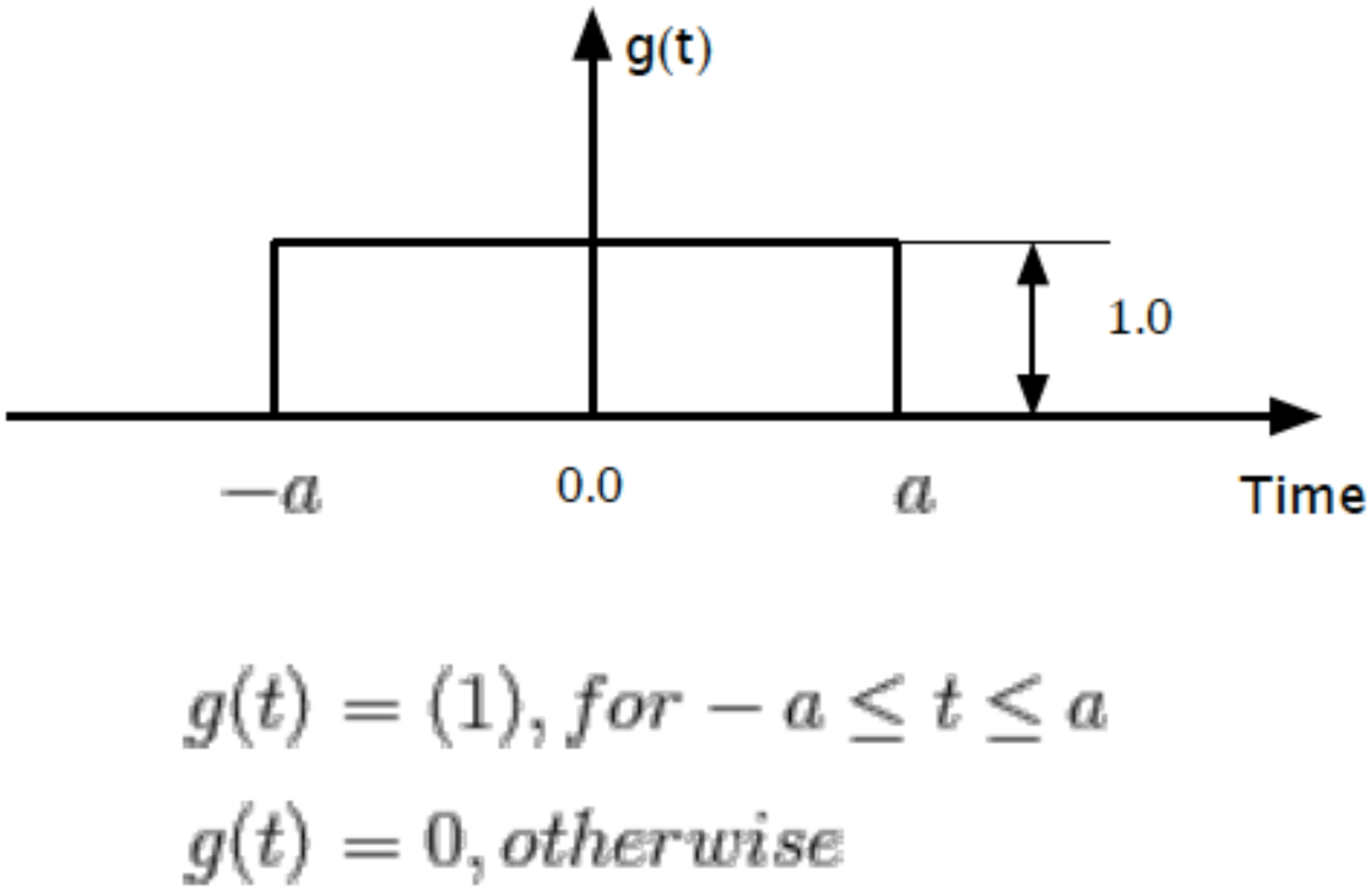}
  \caption{We consider an arbitrary function.}
  \label{fig12}
  \end{center}
\end{figure}

% equation (37)
\begin{equation}
F(\Omega )=\int_{-\infty }^{\infty }g(t)\cdot exp(-i\Omega t)\cdot dt=\int_{-a}^{a}g(t)\cdot exp(-i\Omega t)\cdot dt=\frac{2}{\Omega }\cdot sin(a\Omega ),
\end{equation}
where $\Omega$ is given by,  \\
\[ \Omega =\frac{2\pi }{\Delta t}=2\pi f(frequency). \]   \\
We plot the obtained result of the equation (31) with respect to the value of "$  a  $" and show the frequency spectra in Fig.13. Depending on the smaller values of "$  a  $" in the equation of (31), the peak spectrum becomes wider and wider. This contains an important physical meaning. If an arcing occurs in a short time and it is expressed by delta function, the delta function is transferred by Fourier transform and obtained frequency spectrum becomes flat. It means that power levels become constant at any frequency points. Therefore delta function is used to express white noise. As an example, the lightning occurs in a short time of 1$\mu$s, and expected peak frequency is obtained f=1/$\Delta$t = 1MHz. It means that natural lightning gives lower frequency spectrum. Thus if one detects different frequency points at the same time, one could separate air shower events from natural or artificial noises. 
On the other hand, air shower event has different frequency spectrum as shown in Fig.2. The frequency is continuous. This is the reason why electrons and positrons produced by cascade shower have various energies, and the obtained frequency spectrum superposes various frequency spectra and results in continuous frequency spectrum.  

In order to separate air shower events from various noises, it could be better to measure both polarization and frequency spectrum at the same time.

%
% Figure 13 should be inserted here
% Fig_13
\begin{figure}
\begin{center}
  \includegraphics[scale=0.4]{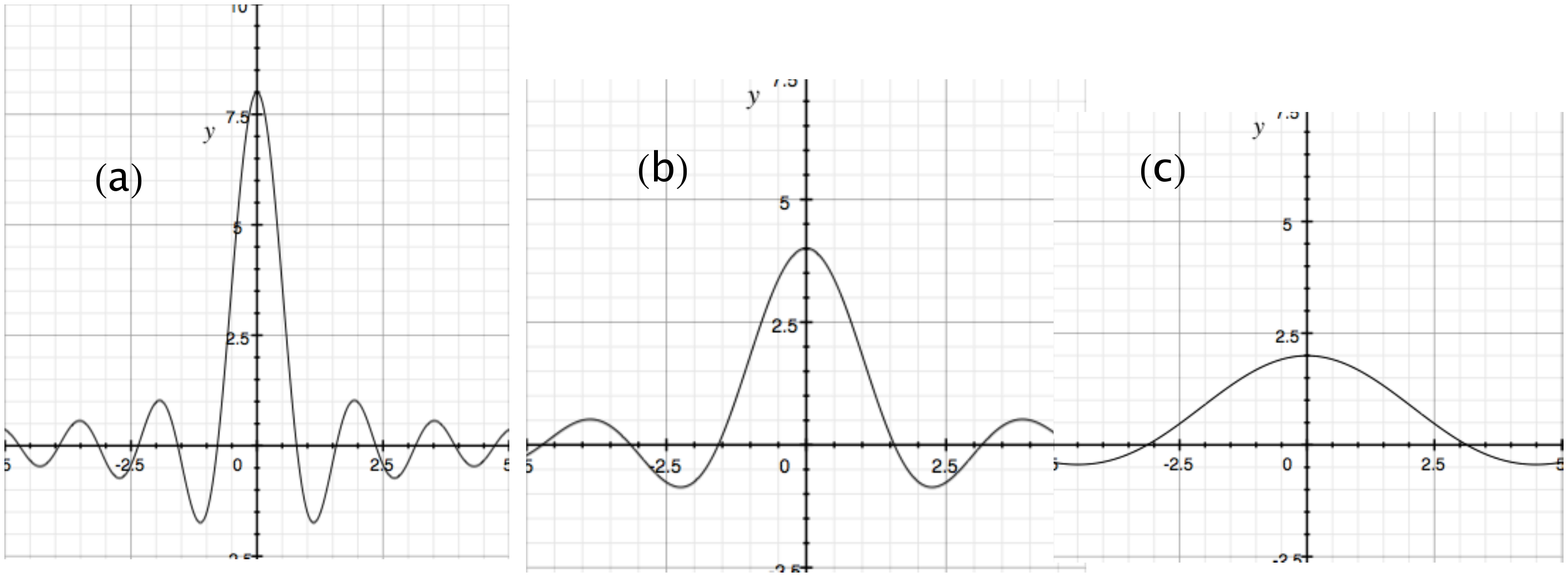}
  \caption{A signal with rectangular shape is transferred to frequency domain by using Fourier transform, (a): the case of  $a$ =4, (b):  $a$=2 and (c): $a$ =1, respectively.}
  \label{fig13}
  \end{center}
\end{figure}

 %% Section 9
\section{ Radio emission power from air shower events}
%section 9.1
\subsection{\bf Relation between radio emission power and primary cosmic ray energy}
Frequency spectrum as shown in Fig.2 suggests that we should detect lower frequency region than 100 MHz. In particular, RF power level in higher frequency region rapidly decreases with the distance between an antenna and the center of air shower events. Since the wavelength in high frequency region becomes shorter than the disk length of electrons and positrons as shown in Fig.1, the coherency of synchrotron radiation does not occur anymore in lower frequency region. Thus the coherency can be expected only in lower frequency region of 10 MHz, since the duration of air shower is around 100 ns and it is converted into the frequency of 10 MHz \cite{Huege}. 

According to Huege \cite{Huege}, RF power intensity of radio emission is proportional to $ E^{0.96} $ for incident primary particle energy "E". In actual, we may assume that the expected radio emission power is proportional to primary cosmic ray energy linearly. It means that electro-magnetic field strength as shown in Fig.2 increases linearly with primary cosmic ray energy. In particular, lower frequency region than 100 MHz becomes easier to detect radio emission.

%section 9.2
\subsection{\bf Radio emission power at 20 m and 140 m points}
Let us discuss RF detection at 20 m and 140 m points. In particular, we are interesting in the power level at 140 m point, since we can save the total number of antenna detectors. The numerical RF power levels from 10 MHz to 100 MHz based on Huege \cite{Huege} are listed in Table 6. The primary cosmic ray energy is fixed at $ 1.0 \times 10^{17}$ eV. The antenna factors as shown in Fig.3 or Table 2 are not available anymore. We need to estimate antenna factors in the lower frequency region. To estimate antenna factor, the formula in the equation of (36) in Appendix A is available. We have to do some assumptions for the calculation of antenna factor. In the equation of (36), we neglect the parts of polarization and impedance matching. Thus we obtain a simple equation,
% equation # (27)
\begin{equation}
AF(dB/m)=20log[f(MHz)]-10log(G(dB))-10log(R_{L})-12.8.
\end{equation}
In the right side of equation, we may take the input impedance 50 ohms. In order to obtain the gain, we adopt both axial-mode helix and Yagi-Uda antennas. To obtain the value of gain, the equation (30) is available. Since the value of k is not known, we have to assume the value of k. According to a textbook \cite{Stutzman}, if an antenna is manufactured precisely, the value of k becomes 1. So we assume the k value 0.5 and the gain is presented.
% equation # (28)
\begin{equation}
G=k\cdot D=0.5\cdot D.
\end{equation}
The values of directivity D have already calculated and shown in Table 4 and 5. Thus the equation (32) is rewritten by
% equation # (29)
\begin{equation}
AF(dB/m)=20log(f(MHz)-10log(0.5\cdot D)-10log(50)-12.8=20log(f(MHz)-10log(0.5\cdot D)-29.79.
\end{equation}
We input D=18 for axial-mode helix antenna from Table 4 and also D=9.1 for Yagi-Uda antenna from Table 5. We can calculate antenna factors for both antennas, and results are summarized in Table 7. 

%% Table 6
\begin{table}[ht]
\begin{center}
\begin{footnotesize}
\caption{Numerical electric field strengths from 10 MHz to 100 MHz are shown in case of $ E=1.0 \times 10^{17} $ eV  \cite{Huege}.}
%\begin{tabular}{llr}
%\begin{tabular}{lcc}
% Table # 6
\begin{tabular}{|c| l| l| l| l| l| l| l| l| l| l| l| }
\hline
 f  (MHz)  &  10 MHz    &   20 MHz  &   30 MHz  &  40 MHz  &   50 MHz  &  60 MHz  &  70 MHz   &  80 MHz   & 90 MHz   &  100 MHz     \\
 \hline
  20-m point                   &              &               &              &              &              &               &              &             &              &           \\    
  $ (\mu V/m)/(MHz) $ & 10.010  &  8.0731  &  6.5110  &  5.2511 &  4.2350 &  3.4156  & 2.7546  & 2.2216  & 1.7917  &  1.4450  \\
  \hline
   140-m point                    &              &               &              &              &              &               &              &             &              &           \\   
   $ (\mu V/m)/(MHz) $  & 4.6930  & 3.1141 & 2.0664 & 1.3712   & 0.90984 & 0.60373 & 0.40061  &  0.26583  &  0.17639 & 0.11705   \\
 \hline
\end{tabular}
\end{footnotesize}
\end{center}
\end{table}
%% \end{center}

%% Table 7
\begin{table}[ht]
\begin{center}
\begin{footnotesize}
%\begin{tiny}
\caption{ Antenna factors for axial-mode helix and Yagi-Uda antennas as shown in Fig.16 of Appendix C are listed.}
%\begin{tabular}{llr}
%\begin{tabular}{lcc}
% Table # 7
\begin{tabular}{|c| l| l| l| l| l| l| l| l| l| l| l| }
\hline
 f  (MHz)  &  10 MHz  &  20 MHz  & 30 MHz  & 40 MHz  & 50 MHz  &  60 MHz  & 70 MHz  & 80 MHz  & 90 MHz  & 100 MHz     \\
 \hline
  Antenna factor for           &              &               &              &              &              &               &              &             &              &           \\    
 axial-mode helix (dB/m) &- 19.33   & -13.31  &  -9.790  & -7.291   &  -5.353  &  -3.769   & -2.430  &  -1.271  &  0.2476  & 0.6676  \\
  \hline
  Antenna factor for           &              &               &              &              &              &               &              &             &              &           \\   
  Yagi-Uda (dB/m)    &  -18.37 &  -10.35  &  -6.828   & -4.329  &  -2.391  &  -0.8071  &  0.5318  &  1.692  &  2.715     &   3.630   \\
 \hline
\end{tabular}
\end{footnotesize}
%\end{tiny}
\end{center}
\end{table}

%% \end{center}
 In order to estimate RF power level of air shower events in the frequency range from 10 MHz to 100 MHz, we adopt the same configuration as that of Fig.7 and we need to add a band-pass filter with the bandwidth of 10 MHz after the preamplifier. Thus total cable loss and the gain of preamplifier are available for the calculation. We have to substitute the antenna HP11966D to axial-mode helix antenna and Yagi-Uda antenna, respectively. In order to calculate RF power level at the input point as shown in Fig.7, all preparations have done. We carry out the calculation along section 4.1. The obtained results are summarized in Fig.14. Signal levels from 10 MHz to 100 MHz exceed the thermal noise level at the point of 20 m. On the other hand, RF power levels at the point of 140 m are lower than thermal noise level at around 100 MHz frequency regions. However, RF power levels in lower frequency region than around 70 MHz show higher power than thermal noise level. If we could develop a realistic antenna, which means that the size is not large as discussed in section 5 and also covers lower frequency region than 80 MHz, it would be a good detector for radio emission. In particular, to decrease the number of antenna detectors, antennas with the sensitivity in lower frequency regions would become a key component. We so far have been discussed primary cosmic ray energy with $ E=1.0 \times 10^{17} $ eV. If an antenna with a realistic size and the sensitivity of lower frequency regions would be available, we could expand the detection of cosmic ray with the energy of lower than $ E=1.0 \times 10^{17} $ eV. 
 
 It must be useful to mention about RF detector. As one sees in Fig.14, the signal levels from air shower events distribute from -20 dBm to -75 dBm for cosmic ray energy of $ E=1.0 \times 10^{17} $ eV. If primary cosmic ray energy becomes higher, the RF detector with wide dynamic range is required. Therefore a commercialized product of MAX2016 as shown in Fig.10 was introduced as an example.
 
%
% Figure 14 should be inserted here
% Fig_14
\begin{figure}
\begin{center}
  \includegraphics[scale=0.6]{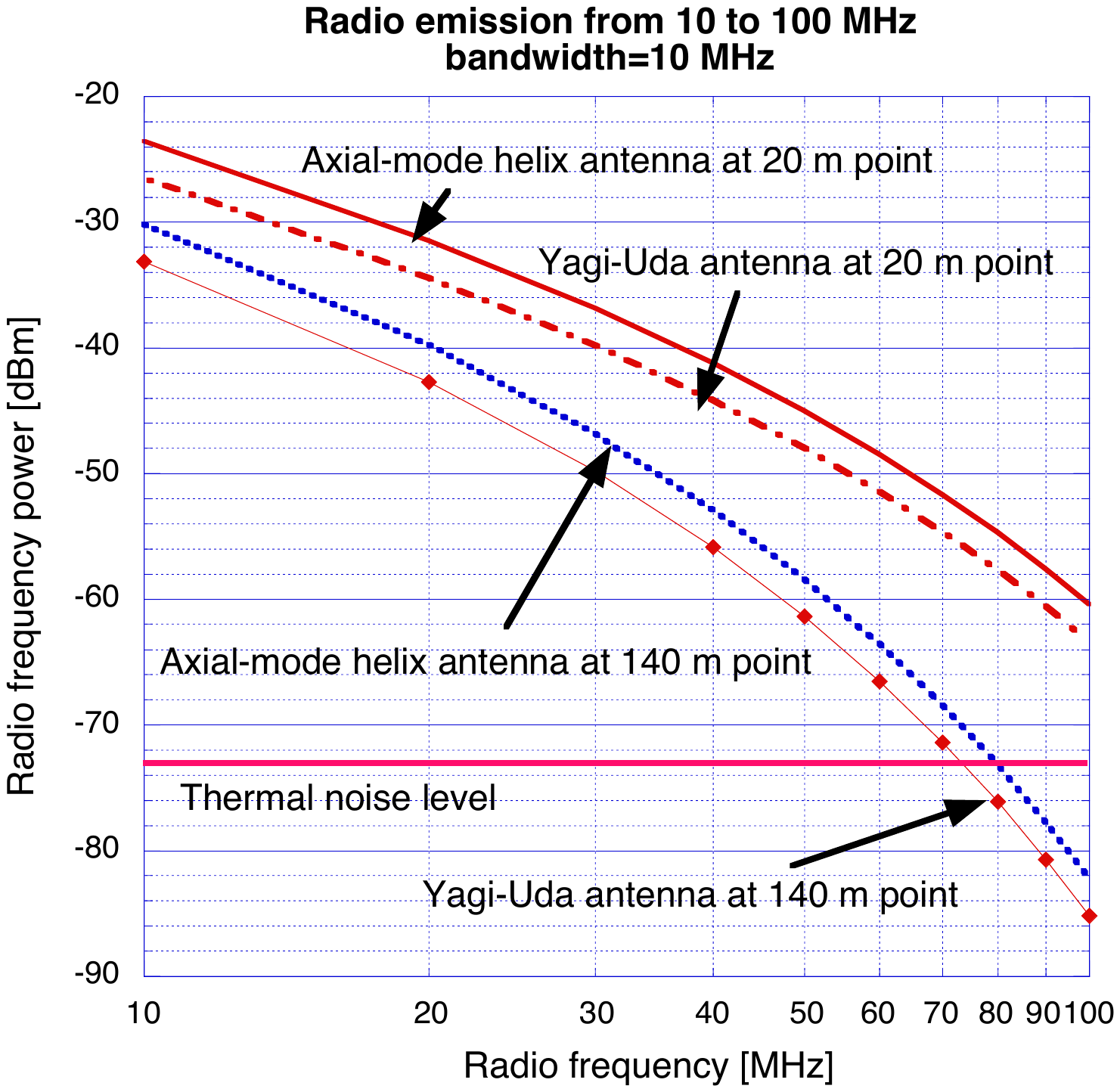}
  \caption{RF power levels at the point of an RF detector as seen in Fig.7 are shown. Radio emission power from cosmic ray event and thermal noise level passing through a 10 MHz band-pass filter are indicated. If radio frequency power level becomes larger than that of thermal noise, one can detect signal from air shower event. Primary cosmic ray energy is $ E=1.0 \times 10^{17}$ eV.}
  \label{fig14}
  \end{center}
\end{figure}

%% Section 10
\section{Summary}
We reviewed the history of detection of radio emission from air shower events. In order to detect radio emission, the selection of antenna is one of the most important factors. A parabolic dish antenna is the best one, however the cost is the most expensive. Therefore we proposed to adopt an axial-mode helix antenna, which has a wide acceptable frequency range. To reduce the number of antenna detectors, antennas with the sensitivity of lower frequency regions are more useful, because one antenna can cover wider area for the detection of radio frequency. We need to develop an antenna with a realistic size and the sensitivity in lower frequency regions than 80 MHz. If we could develop the antenna that we desire, we would be able to detect cosmic ray events even with lower energy than $ 1.0 \times 10^{17}$ eV. We also reduce the number of antenna detectors.

To measure precisely the incident direction and energy of air shower events at the same time, we need to place as many antennas as possible widely on the ground. Those antennas give the information of energy and incident direction of air shower from arrival time differences between antennas and the center of air shower. It is mentioned that a flash-ADC is more useful to measure both arrival time and the power of radio emission at the same time. RF detectors need to have a wider dynamic range to cover wide cosmic ray energy. 

To identify air shower events from natural or artificial noises, we proposed the measurement of frequency spectrum from air shower events. If we could detect the polarization of radio emission as well as frequency spectrum at the same time, it is the best way.

\section*{Acknowledgements}
% \begin{Acknowledgements}
We are indebted to M. Yeddula, who introduced two textbooks about antenna and also discussed with us. If he did not introduce antenna textbooks, this paper could not be seen. We also thank to Y. Ohashi, H. Ego and S. Sasaki for discussion about air shower detection.

 \newpage
% Appendix A
\appendix
\include{99_appendix}
\section{Antenna factor}
\subsection{\bf Antenna factor: definition}
Antenna factor AF is defined as the ratio of the field intensity illuminating $ E^{i}$ to the received voltage $ V_{A}$ across the antenna terminals \cite{Stutzman}:
\begin{equation}
AF=\frac{E^{i}}{V_{A}} (m^{-1}).
\end{equation}
This is an electric field antenna factor. Taking (10 log) of both sides gives \\
$ E_{rms}^{i}(dB\mu V/m)=receiver \   sensitivity=V_{A,rms}(dB\mu V)+AF(dB/m), $ \\
where
\begin{equation}
AF(dB/m)=20log(f(MHz))-10log(G(dB))-10log(R_{L})-10log(p)-10log(q)-12.8.
\end{equation}
\hspace{3cm} $ R_{L}=50 \Omega $,   \\
\hspace*{3cm} G : gain,   \\
\hspace*{3cm} $p : polarization $,  \\
\hspace*{3cm} $ q : impedance \  match $.  \\
The antenna factor is included as one of the basic parameters of an antenna, as it is used almost exclusively in electro-magnetic interference (EMI) testing when making radiated electric field strength (E-field) measurements. Electric field measurements are necessary for determining compliance with most electromagnetic interference requirements/standards. The antenna factors vary with respect to radio frequency, so the appropriate antenna factor must be applied, based on the radio frequency of the signal being measured.

Let us show an example. An antenna detects electric-field strength as shown in Fig.15. 

In practice, to determine an actual electric-field strength level E, we need to make correction:
\begin{equation}
 E=R_{receiver}-P_{gain}+L_{cable}+AF, 
\end{equation}
\hspace{3cm}  $R_{receiver}$ : receiver  reading,\\
\hspace*{3cm} $P_{gain}$ : pre-amplifier  gain,\\
\hspace*{3cm} $L_{cable}$ : cable  loss,,\\
\hspace*{3cm} $AF : antenna$  \  factor. \\

If A $dB \mu V$ is measured at the antenna output connector, the gain B dB of pre-amplifier, cable loss C dB and the antenna factor at the frequency of measurement is D dB/m, the value of the electric-field is determined as follows:
\begin{equation}
E=A(dB\mu V: receiver)-B(dB:gain)+C(dB:loss)+D(dB/m:AF).
\end{equation}
If we have the value of antenna factor, we can know the input radio emission power correctly.

 %
% Figure 15 should be inserted here
% 
\begin{figure}
\begin{center}
  \includegraphics[scale=0.3]{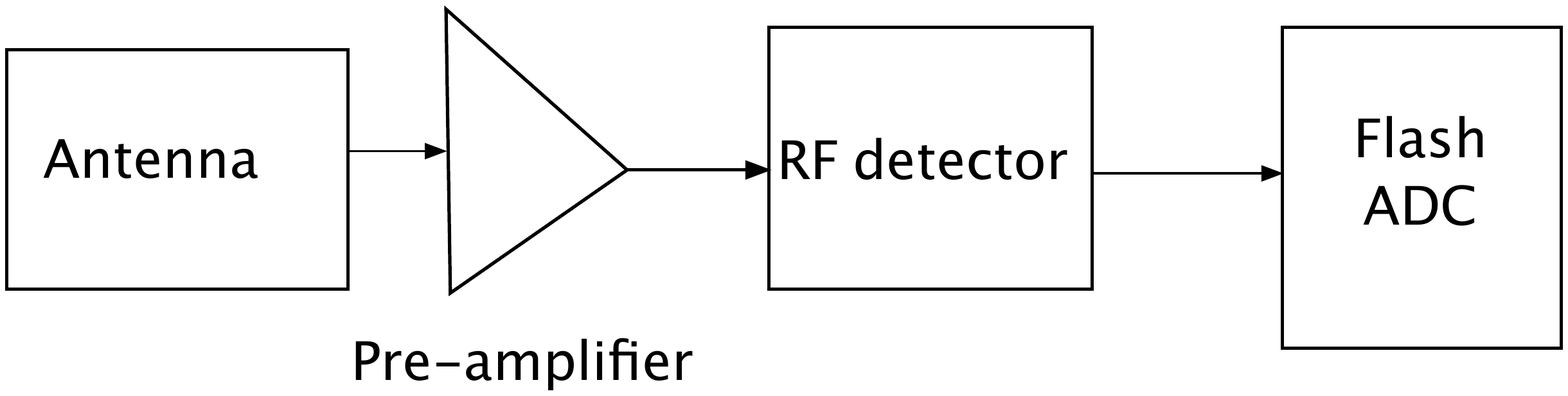}
  \caption{  Electric field detection.}
  \label{fig15}
  \end{center}
\end{figure}

\subsection{\bf Calculation}
 In order to obtain antenna factor, calculation is available. The formula is given in (36). Calculation error using the formula (36) becomes small at around 100 MHz and lower frequency region, because wavelength at low frequency region is so long that antenna factor is not largely dependent on the accuracy of antenna structure.

% Appendix B
\newpage
\section{Various Units related to radio frequency (RF)}
Let us show various units about RF.  \\
 % definition 1
{\bf  Definition 1} \\
 It is common to handle RF power in the unit of dBm. In general A mW (mili-watts) power is written by
 \begin{equation}
 10log(A)=(B)dBm.
 \end{equation}
As an example, let us show 1 W (=watt: power unit) by dBm unit. 1 W is equal to 1000 mW. The dBm unit handles mW and it is converted by using the definition of (39),
 \begin{equation}
1W= 10log(1.0\times 10^{3})=30dBm.
 \end{equation}
Now, let us try to the inverse calculation. How does one convert 0 dBm to the unit of mW? Using the definition (39), 0 dBm is expressed,
 \begin{equation}
 10log(A)=0dBm, \ \  0^{0}=1mW, \ \ \therefore 0dBm=1mW.
 \end{equation}
  % definition 2
{\bf  Definition 2} \\
 An antenna detects electric field. Therefore following unit is useful,
  \begin{equation}
 \mu V/m ,
 \end{equation}
 where V stands for voltage and  "$  m  $" means length in the metric system. Next unit is also useful;
 \begin{equation}
  dB\mu V/m.
 \end{equation}
  As an example,
  \begin{equation}
  100\mu V/m=40dB\mu V/m.
 \end{equation}
  (44) is calculated as following:
  \begin{equation}
  20log100=40.
 \end{equation}
 % definition 3
 {\bf  Definition 3} \\
 RF power P has a relation to voltage V. If input impedance is R ohm ($ \Omega $), in case of accelerator it is called shunt impedance, 
  \begin{equation}
  P=\frac{V^{2}}{R}.
 \end{equation}
  As an example, the voltage of $ 0 dB\mu V$ is received with 50 $ \Omega $ impedance. The $ 0 dB\mu V$ is expressed by voltage unit,
  \begin{equation}
 0dB\mu V=1.0\times 10^{-6}volt. 
  \end{equation}
One inputs the values of voltage and impedance into (46),  
  \begin{equation}
  P=\frac{(1.0\times 10^{-6})^{2}}{50}(watts)=1000\times \frac{(1.0\times 10^{-6})^{2}}{50}(mW).
 \end{equation}
  One inputs (48) into (39) and obtains the following relation,
   \begin{equation}
  0dB\mu V=10\times log(1000\times \frac{(1.0\times 10^{-6})^{2}}{50})=-106.99=-107dBm.
 \end{equation} 
 Obtained relation of (49) is useful to convert from electric field unit to power unit.

 % Appendix C
 \newpage
\section{Various antennas \cite{Kraus}}
Four antennas as shown in Fig.16: Parabolic dish antenna is used for radio communication and radio astronomy. Yagi-Uda antenna is the most popular for a receiver of television. Axial-mode helix antenna is useful for astronomy. Since the rectangular horn antenna is strong enough for impact, it is useful to learn antenna technology. 
  %
% Figure C.16 should be inserted here
% 
\begin{figure}
\begin{center}
  \includegraphics[scale=0.5]{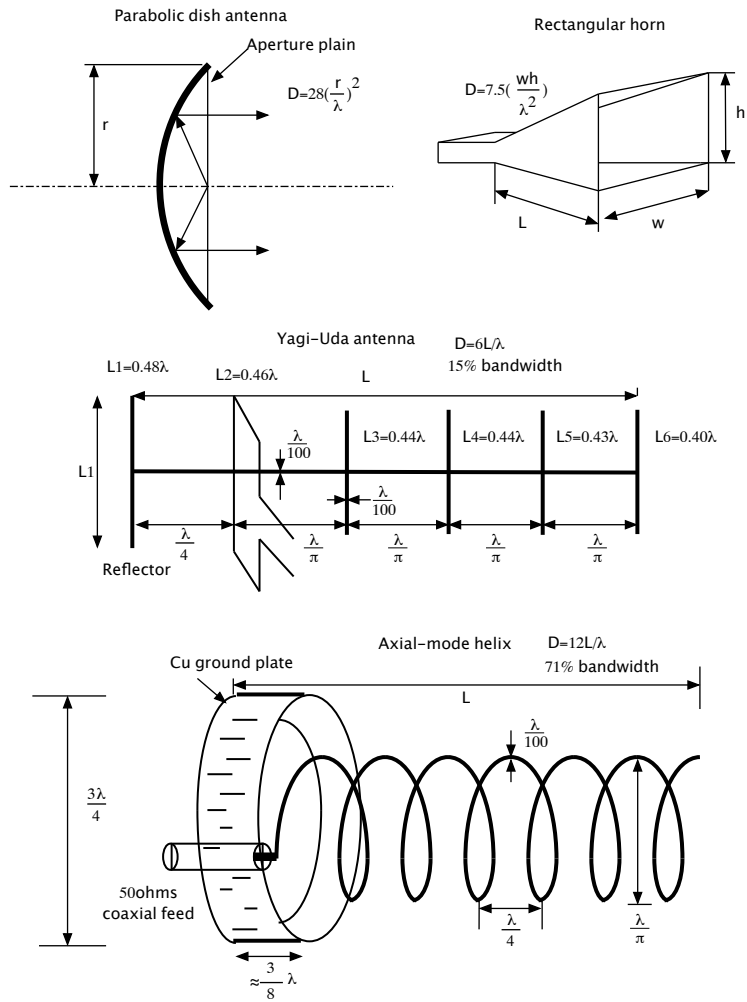}
  \caption{Various antennas and related parameters}
  \label{fig16}
  \end{center}
\end{figure}

\end{document}